# Minimum-Complexity Failure Correction in Linear Arrays via Compressive Processing

F. Zardi, G. Oliveri, *Senior Member, IEEE*, M. Salucci, *Member, IEEE*, and A. Massa, *Fellow, IEEE*

***Abstract*—Given an array with defective elements, failure correction (*FC*) aims at finding a new set of weights for the working elements so that the properties of the original pattern can be recovered. Unlike several *FC* techniques available in the literature, which update all the working excitations, the Minimum-Complexity Failure Correction (*MCFC*) problem is addressed in this paper. By properly reformulating the *FC* problem, the minimum number of corrections of the whole excitations of the array is determined by means of an innovative Compressive Processing (*CP*) technique in order to afford a pattern as close as possible to the original one (i.e., the array without failures). Selected examples, from a wide set of numerical test cases, are discussed to assess the effectiveness of the proposed approach as well as to compare its performance with other competitive state-of-the-art techniques in terms of both pattern features and number of corrections.**

## I. Introduction

RECENTLY, there has been an increased interest in phased arrays comprising hundreds of elements for several applications ranging from in-flight connectivity up to advanced radar technology (see [1]-[5] and the reference therein). Indeed, the cost and the complexity of array systems have reduced thanks to the availability on the market of transmit and receive modules transmit and received modules [6] and the development of simplified feeding networks [7][8]. Furthermore, more and more challenging requirements in many and relevant civil, commercial, and military applications have further stimulated/forced the adoption of large array systems. For instance, in mobile communications, the massive MIMO paradigm is driving the request of more and more antennas built in next-generation mobile phones and base stations [9][10]. On the other hand, phased arrays with large apertures are investigated in satellite systems to counteract the heavier path loss at higher frequency bands [11][12]. Moreover, phased arrays with thousands of elements are going to replace the previous generation of highly-directive parabolic reflectors in radar meteorology [13][14]. Of course, as the number of transmit and receive modules grows, the probability that a failure occurs also increases and suitable countermeasures need to be envisaged to prevent the loss of working functionality of the system as well as to guarantee its reliability.

Failure correction (*FC*) aims at reconfiguring the working elements of a faulty array to recover (all or the key) pattern features afforded by the original whole array to guarantee consistent performance of the system. Towards this end, many correction techniques, which differ in the problem formulation and/or the solution strategy, have been proposed. First, correction methods involving efficient numerical implementations that require few computational resources have been developed. As an example, the *FC* problem has been formulated in [15] as a minimization one with quadratic constraints, then solved with a fast method devoted to minimize the average power within the sidelobe region. Similarly, a correction method based on conjugate gradients that reduces the average sidelobe provided an increment of the mainlobe beamwidth has been derived in [16]. In [17], the linear least square deviation from a reference pattern has been minimized.

Although these approaches give *FC* solutions in a fast and efficient way, they are not suitable for directly minimizing high-level pattern features such as the Side Lobe Level (*SLL*) or the directivity. In order to address such a challenge, a category of *FC* methods based on alternating projection methods and initially devised for the synthesis of large arrays is available in the state-of-the-art literature. In [18], the Vector-Space Projection algorithm has been employed to jointly optimize different array characteristics concerned with both the radiated pattern (e.g., the *SLL* and the total transmitted power) and the array architecture (e.g., the maximum excitation magnitude). Moreover, Keizer exploited in [19] the invertible Fourier-based relation between the array excitations and the corresponding array factor to compute the corrected weights to recover a reference pattern. In [20], the alternate $\ell_2$-norm projection method has been combined with a sparse failure detection strategy to correct a failed planar array with $N = 289$ elements.

Reviewing the *FC* literature, it is worth mentioning the use of global optimization techniques based on Genetic Algorithms (*GA*s). For instance, Yeo *et al.* analyzed different mating schemes and reported numerical results for two- and three-

Manuscript received May 0, 2020

This work has been partially supported by the Italian Ministry of Education, University, and Research within the Program Smart cities and communities and Social Innovation (CUP: E44G14000060008) for the Project WATERTECH - Smart Community per lo Sviluppo e l'Applicazione di Tecnologie di Monitoraggio Innovative per le Reti di Distribuzione Idrica negli usi idropotabili ed agricoli (Grant no. SCN_00489) and within the Program PRIN 2017 for the Project Cloaking Metasurfaces for a New Generation of Intelligent Antenna Systems (MANTLES).

F. Zardi, G. Oliveri, M. Salucci, and A. Massa are with the ELEDIA@UniTN (DISI - University of Trento), Via Sommarive 9, 38123 Trento - Italy (e-mail: {francesco.zardi, giacomo.oliveri, marco.salucci, andrea.massa}@unitn.it)

G. Oliveri and A. Massa are also with the ELEDIA Research Center (ELEDIA@L2S - UMR 8506), 3 rue Joliot Curie, 91192 Gif-sur-Yvette - France (e-mail: {giacomo.oliveri, andrea.massa}@l2s.centralesupelec.fr)

A. Massa is also with the ELEDIA Research Center (ELEDIA@UESTC - UESTC), School of Electronic Engineering, Chengdu 611731 - China (e-mail: andrea.massa@uestc.edu.cn

A. Massa is also with the ELEDIA Research Center (ELEDIA@TSINGHUA - Tsinghua University), 30 Shuangqing Rd, 100084 Haidian, Beijing - China (e-mail: andrea.massa@tsinghua.edu.cn

element failures [21]. In [22], a GA-based *FC* approach has been proposed jointly with an adaptive weighted pattern mask aimed at updating, throughout the optimization process, the pattern constraints along user-defined angular regions. The optimization of the *SLL*, the directivity, and the Dynamic Range (*DR*) of the array excitations as well as the number of corrections has been dealt with in [23] by means of the *GA*-based minimization of a single-objective cost function. In [24], *GA*s have been still applied to synthesize the array architecture that minimizes the failure-probability-weighted pattern deviation from the whole/original one due to the failure of each element of the array. Such an approach provides a viable solution for arrays without reconfiguration capabilities, but deteriorations of the radiated pattern may arise during normal operations.

Finally, the recovery of the signals not received at the faulty elements of an array has been addressed in [25][26] by assuming the *a-priori* (although not precise) knowledge of the directions-of-arrival of plane waves impinging on the array.

As a general comment on *FC* techniques, it is worthwhile to point out that the complete restoration of the original pattern is generally not possible and, almost always, a trade-off solution in terms of pattern features and array performance is looked for. Furthermore, most of the state-of-the-art solutions result in many - or even all - working elements being reconfigured. However, while the modification of the element excitations can be done on-line for reconfigurable arrays, it might be desirable in other array architectures to minimize the number of corrections so that reducing the maintenance costs and the system down-time. Additionally, both the availability and the number of spare parts can be limited in some applications (e.g., the on-orbit satellite servicing [27][28]).

Within the *FC* framework, this work addresses the problem of finding the minimum set of corrections for the restoration of a user-defined performance metric in a faulty array. By formulating the Minimum-Complexity Failure Correction (*MCFC*) problem as a non-deterministic polynomial-time hard $\ell_0$-norm minimization one with non-linear constraints, a correction method based on the Compressive Processing (*CP*) paradigm (see [29]-[31] and the reference therein) is then developed. The proposed approach combines a $\ell_1$-norm relaxation of the $\ell_0$-norm and a backtracking strategy for an efficient sampling of the high-dimensional solution space to yield a satisfactory correction of the array excitations in a proper amount of time. The key motivations of these choices can be summarized as follows. First, the practice of approximating the $\ell_0$-norm with the $\ell_1$-norm is often adopted when exploiting the *CP* paradigm [32][33]. Indeed, the arising approximated cost function turns out to be convex so that it can be minimized with any of the many available implementations of convex-optimization methods. Moreover, theoretical analyses have shown that, under specific conditions, the $\ell_1$-norm minimization is equivalent to the $\ell_0$-norm one [34]-[36]. As for the backtracking algorithm, it has proven reliable in solving *CP* problems [37][38]. As a matter of fact, it has been successfully applied in electromagnetics to the sparse reconstruction of scatterers in through-the-wall imaging [39] and to the optimal signal sampling for bandwidth enhancement in phased arrays [40].

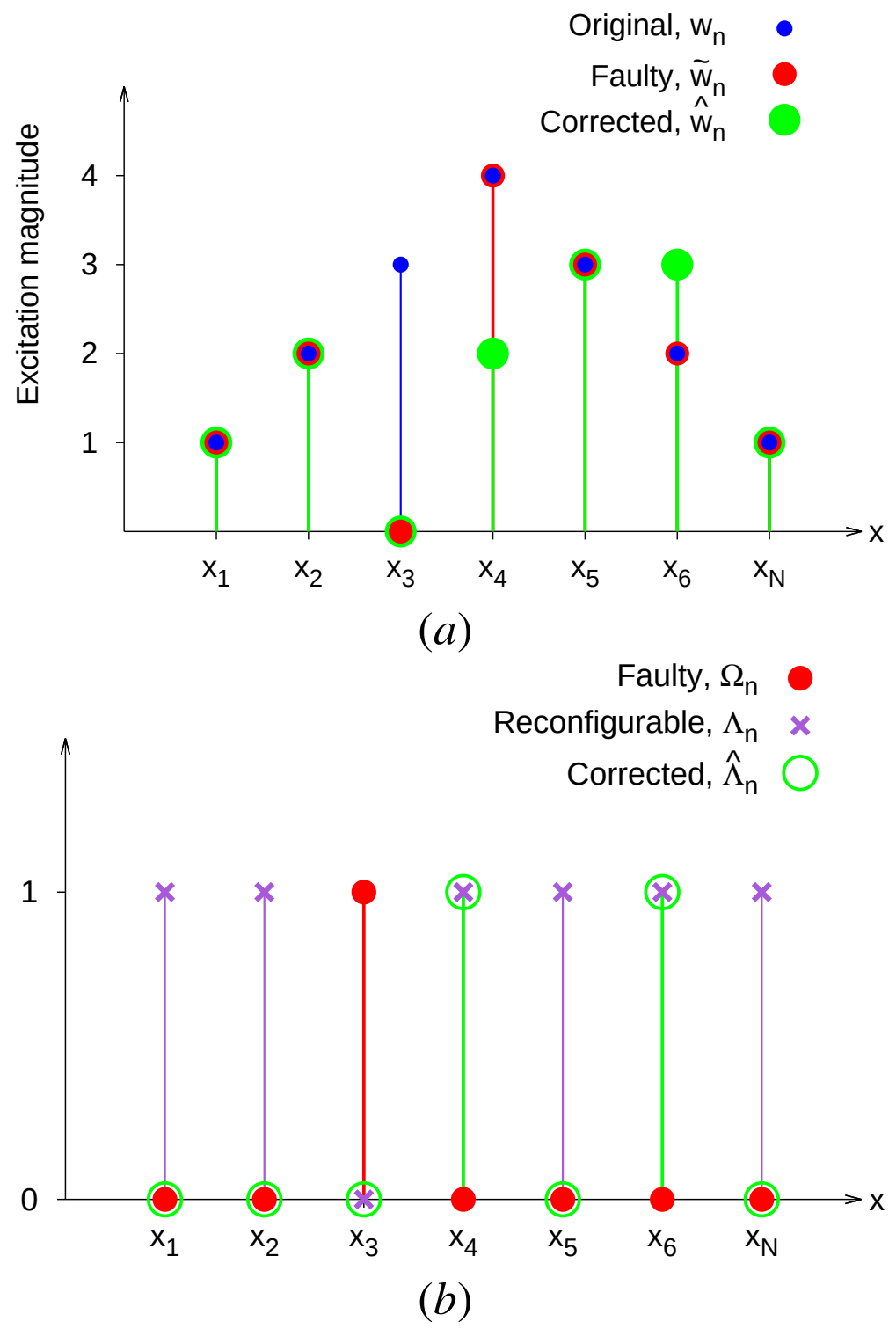

Figure 1. *Illustrative Example* ($N = 7$, $N_F = 1 \rightarrow \eta_F = 14$ %) - Pictorial representation of the excitation set of the "*original*" array, $\mathbf{w} = \{1, 2, 3, 4, 3, 2, 1\}$, of the "*damaged*" array, $\widetilde{\mathbf{w}} = \{1, 2, 0, 4, 3, 2, 1\}$, and of the "*corrected*" array, $\widehat{\mathbf{w}} = \{1, 2, 0, 2, 3, 3, 1\}$, along with the corresponding binary vectors indicating the locations of the failed elements, $\mathbf{\Omega} = \{0, 0, 1, 0, 0, 0, 0\}$, of the reconfigurable/admissible elements, $\mathbf{\Lambda} = \{1, 1, 0, 1, 1, 1, 1\}$ ($\mathbf{\Lambda} \triangleq \mathbf{1} - \mathbf{\Omega}$), and of the corrected elements, $\widehat{\mathbf{\Lambda}} = \{0, 0, 0, 1, 0, 1, 0\}$.

The main contributions of this paper are: (*a*) a formulation of the *MCFC* problem that allows one to adopt arbitrary metrics for dealing with any user-chosen array requirements; (*b*) the definition of a mathematical/theoretical framework suitable for an effective and reliable application of *CP*-based techniques to minimize the number of excitations to be reconfigured for recovering the pattern features of the original array; (*c*) the introduction of an innovative *CP*-based correction method that combines a $\ell_1$-norm relaxation of the $\ell_0$-norm with a backtracking strategy.

The outline of the paper is as follows. The *MCFC* problem is stated and mathematically formulated in Sect. II. Section III details the procedural steps of the proposed correction method, while Section IV reports some representative numerical results to assess the arising correction performance in a comparative study, as well. Concluding remarks are finally drawn (Sect. V).

## II. Problem Statement

A linear array comprising $N$ isotropic elements, which are distributed along the $x$ axis at the positions $\{x_n;\, n = 1, ..., N\}$ [Fig. 1(*a*)], affords the following radiation pattern

$$F(u) = \sum_{n=1}^{N} w_n \psi_n(u) \tag{1}$$

where $\psi_n(u) = e^{jkx_n u}$ $[u = \sin(\theta)]$ and $k$ is the wavenumber ($k = \frac{2\pi}{\lambda}$), while $\mathbf{w}$ is the set of excitations, $\mathbf{w} = \{w_n;\ n = 1, ..., N\}$, $w_n$ being the working excitation of the $n$-th element of the "*original* " array (i.e., the whole array without failures). Let us assume that $N_F$ radiating elements, whose locations are denoted by the non-null entries of the $N$-elements binary vector $\boldsymbol{\Omega}$ [i.e., $\Omega_n = 1$ if the $n$-th element of the original array is faulty and $\Omega_n = 0$ otherwise - Fig. 1($b$)], are damaged and they are not reconfigurable anymore. Accordingly, the $n$-th ($n = 1, ..., N$) element of the excitation vector of the damaged array, $\widetilde{\mathbf{w}}$ [Fig. 1($a$)], turns out to be

$$\widetilde{w}_n = w_n (1 - \Omega_n), \quad (2)$$

while the radiated pattern is given by

$$\widetilde{F}(u) \triangleq \sum_{n=1}^{N} \widetilde{w}_n \psi_n(u). \quad (3)$$

Although $N_F$ excitations of the faulty array cannot be modified, the remaining $N_C$ ($N_C \triangleq N - N_F$) elements can still be reconfigured to perform the array correction [Fig. 1($b$)]. By indicating with $\widehat{\mathbf{w}}$ [$\widehat{\mathbf{w}} = \{\widehat{w}_n;\ n = 1, ..., N\}$ - Fig. 1($a$)] the set of array excitations after correction, the associated radiated field is

$$\widehat{F}(u) = \sum_{n=1}^{N} \widehat{w}_n \psi_n(u) \quad (4)$$

the weights of the $N_F$ failed elements being set to zero (i.e., $\widehat{w}_n \Omega_n = 0$), while the corrections of the working elements are coded into the excitation correction vector $\Delta\mathbf{w}$

$$\Delta\mathbf{w} \triangleq \widehat{\mathbf{w}} - \widetilde{\mathbf{w}}. \quad (5)$$

The *FC* problem at hand can be then formalized with the following statement

*Minimum-Complexity Failure Correction* (*MCFC*) *Problem* - Given the "*original* array" excitations set, $\mathbf{w}$, the set of $N_F$ faulty elements, $\boldsymbol{\Omega}$, and the excitations of the "*faulty* array", $\widetilde{\mathbf{w}}$, find the optimal correction set $\Delta\mathbf{w}_{opt}$ so that

$$\Delta\mathbf{w}_{opt} = \arg\min_{\Delta\mathbf{w}} \left\{ \|\Delta\mathbf{w}\|_0 \left| \Phi\left(\Delta\mathbf{w}^{(k)}\right) \right. \le \Phi_{target} \text{ and } \Delta w_n \Omega_n = 0 \right\} \quad (6)$$

where $|$ stands for "subject to", $\|\cdot\|_0$ denotes the $\ell_0$-norm, $\|\Delta\mathbf{w}\|_0$ is the number of corrections ($\widehat{N_C} \equiv \|\Delta\mathbf{w}\|_0$), and $\Phi(\Delta\mathbf{w})$ is a single function measuring the array performance when $\Delta\mathbf{w}$ is applied, while $\Phi_{target}$ is the fixed user-defined target value of the array performance for a reliable working of the radiating system.

It is worth pointing out that the choice of $\Phi(\Delta\mathbf{w})$ depends on the applicative context and it can take into account classical array parameters (e.g., *HPBW*, *SLL*, and Directivity) or high-level system requirements such as the link budget or capacity. For instance, $\Phi(\Delta\mathbf{w}) \triangleq HPBW(\Delta\mathbf{w})$ and $\Phi_{target} \triangleq HPBW_{original}$ can be considered if the beamwidth of the original array $HPBW_{original}$ is the fundamental feature

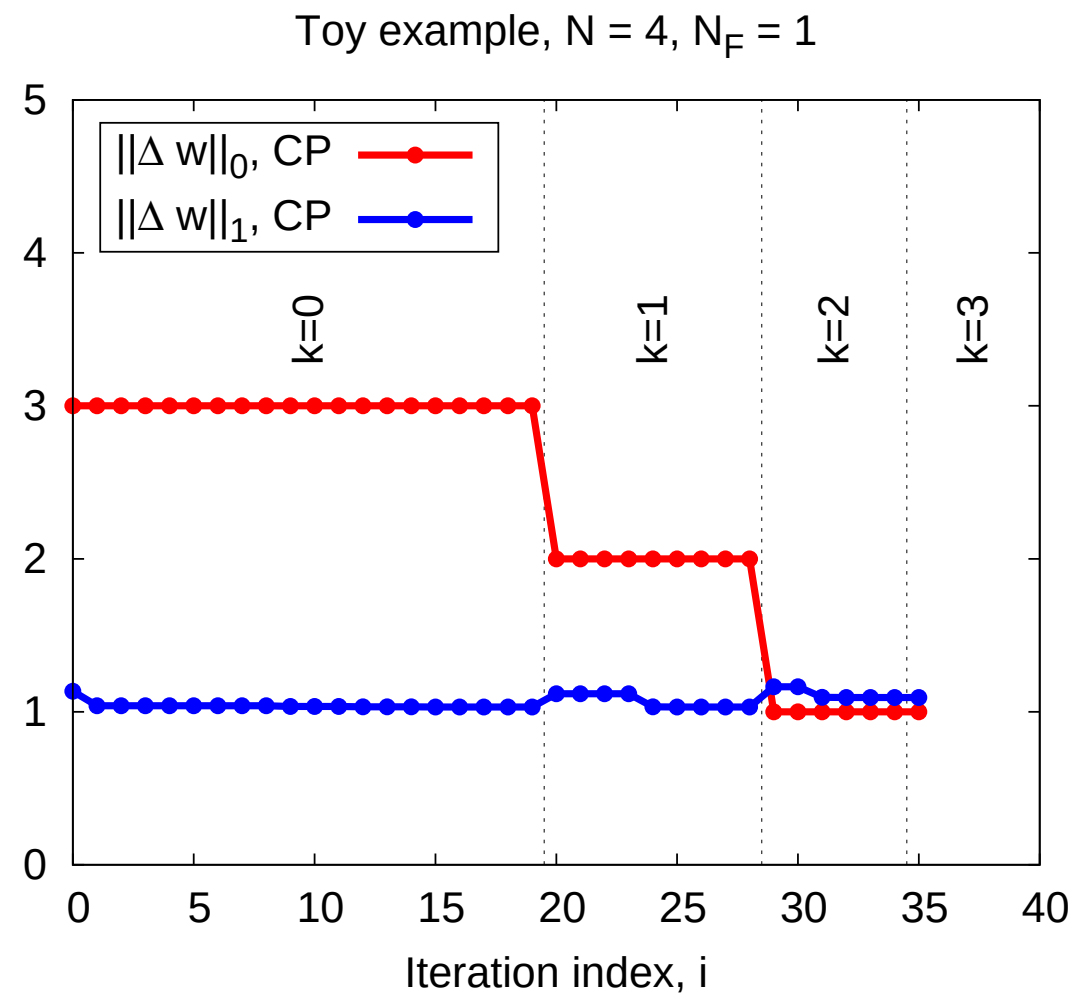

($a$)

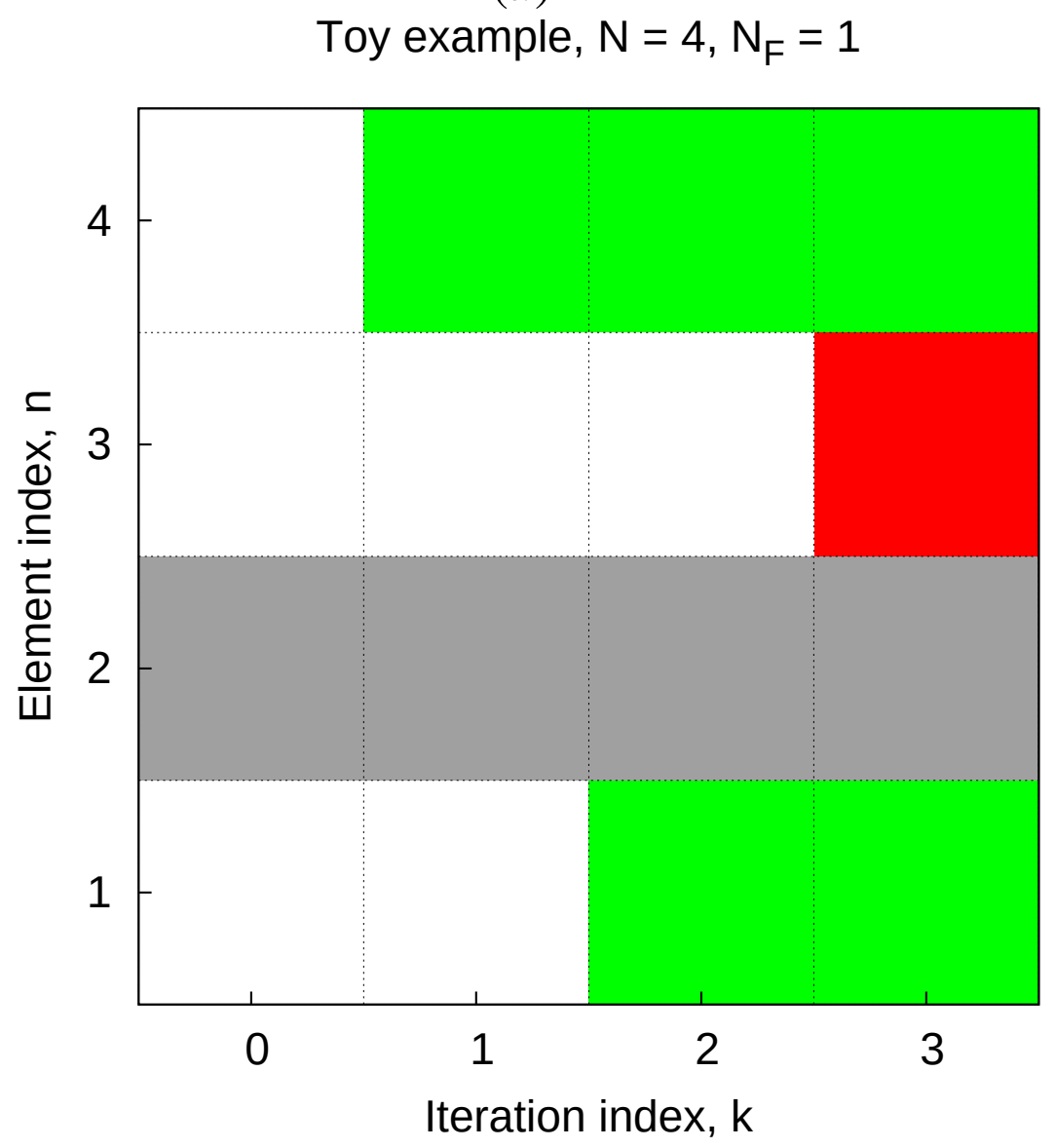

($b$)

Figure 2. *Illustrative Example* ("*Toy Example*": $N = 4$, $N_F = 1 \rightarrow \eta_F = 25$ %; *DC SLL* $= -15$ [dB], $SLL_{target} = -5.5$ [dB]) - Evolution of ($a$) the $\ell_1$-norm and the $\ell_0$-norm of the excitation correction vector, $\Delta\mathbf{w}$, and of ($b$) the status ("required/non-required") of the element corrections, $r_n/s_n$ ($n = 1, ..., N$) ["required" correction ($r_n = 1$) $\rightarrow$ red; "non-required" correction ($s_n = 1$) $\rightarrow$ green; failed element ($\Omega_n = 1$) $\rightarrow$ gray; "unknown"-status correction ($r_n = s_n = \Omega_n = 0$) $\rightarrow$ white], versus the *CP* iteration index, $k$.

to be recovered. Furthermore, the formulation in (6) can be seamlessly extended to include multiple array characteristics into the algorithm by recasting $\boldsymbol{\Phi}(\Delta\mathbf{w})$ and $\boldsymbol{\Phi}_{target}$ to vectorial quantities.

## III. *CP*-Based Solution Technique

Before presenting the *CP*-based correction method, let us point out that the direct $\ell_0$-norm minimization of the corrected weights as in (6) is a non-deterministic polynomial-time hard problem [41]. Indeed, finding the optimal solution, $\Delta\mathbf{w}_{opt}$, would require testing all possible combinations of $\widehat{N_C}$ ($\widehat{N_C} = 1, ..., N_C$) corrections, whose number grows exponentially with $N$. On the other hand, one needs to note that, if the optimal set of correction positions [i.e., the binary vector $\widehat{\Lambda} = \left\{\widehat{\Lambda}_n;\, n = 1, ..., N\right\}$ where $\widehat{\Lambda}_n = 1$ if the $n$-th array element is corrected and $\widehat{\Lambda}_n = 0$, otherwise - Fig. 1($b$)] was *a-priori* known, then the original failure correction problem (6) would reduce to a much simpler synthesis problem aimed at determining the weights of the corrected excitations, $\widehat{\mathbf{w}}$. Moreover, the proposed approach also takes advantage from the fact that, within the framework of $\ell_0$-norm minimization problems, the backtracking algorithm has yielded remarkable results [37][38].

Table I
*Illustrative Example* ("*Toy Example*": $N = 4$, $N_F = 1 \rightarrow \eta_F = 25$ %; *DC* $SLL = -15$ [dB], $SLL_{target} = -5.5$ [dB]) - STEP-BY-STEP DESCRIPTION OF THE EVOLUTION OF THE *CP* VARIABLES THROUGHOUT THE ITERATIVE PROCESS FOR THE CORRECTED-ARRAY SYNTHESIS.

| $k$ | Step | $n_{least}^{(k)}$ | $\mathbf{s}^{(k)}$ | $\mathbf{r}^{(k)}$ | $\Delta\mathbf{w}_{tr}^{(k)}$ | $\Delta\mathbf{w}_{opt}^{(k)}$ | $\left\|\Delta\mathbf{w}_{opt}^{(k)}\right\|_0$ | $\left\|\Delta\mathbf{w}_{opt}^{(k)}\right\|_1$ | $\Phi\left(\Delta\mathbf{w}_{opt}^{(k)}\right)$ |
|---|---|---|---|---|---|---|---|---|---|
| 0 | 0 | $-$ | $\{0,0,0,0\}$ | $\{0,0,0,0\}$ | $-$ | $\{-0.438, 0.0, 0.593, -9.72\times10^{-6}\}$ | 3 | 1.03 | $-5.5$ |
| 1 | 1 | 4 | $\{0,0,0,1\}$ | $-$ | $\{-0.438, 0.0, 0.593, 0.0\}$ | $-$ | $-$ | $-$ | $-$ |
| | 2 | 4 | $\{0,0,0,1\}$ | $-$ | $\{-0.438, 0.0, 0.593, 0.0\}$ | $\{-0.438, 0, 0.593, 0\}$ | 2 | 1.03 | $-5.5$ |
| 2 | 1 | 1 | $\{1,0,0,1\}$ | $-$ | $\{0,0,0.593,0\}$ | $-$ | $-$ | $-$ | $-$ |
| | 2 | 1 | $\{1,0,0,1\}$ | $-$ | $\{0,0,0.593,0\}$ | $\{0,0,1.09,0\}$ | 1 | 1.09 | $-5.5$ |
| 3 | 1 | 3 | $\{1,0,1,1\}$ | $-$ | $\{0,0,0,0\}$ | $-$ | $-$ | $-$ | $-$ |
| | 2 | 3 | $\{1,0,1,1\}$ | $-$ | $\{0,0,0,0\}$ | $\{0,0,0,0\}$ | 0 | 0 | $-2.45$ |
| | 3 | 3 | $\{1,0,0,1\}$ | $\{0,0,1,0\}$ | $\{0,0,0.593,0\}$ | $\{0,0,1.09,0\}$ | 1 | 1.09 | $-5.5$ |
| 4 | 1 | $-$ | $-$ | $-$ | $-$ | $\{0,0,1.09,0\}$ | 1 | 1.09 | $-5.5$ |

In words, the *CP* method starts from a full set of corrections by iteratively reducing its $\ell_0$-norm. At each iteration ($k$ being the iteration index, $k = 1, ..., K$), the algorithm uses a heuristic to guess which correction is of least importance. A trial solution is then generated by removing the least important correction from the current best solution. Subsequently, the $\ell_1$-norm of the trial solution is minimized subject to the pattern requirements. If the requirements are satisfied, the trial solution is stored as the new best solution and the iteration continues until the convergence ($k = k_{opt}$). Otherwise, the algorithm *backtracks* on its previous guess and the correction previously-removed is restored. Towards this purpose, the algorithm keeps track throughout the iterations of which corrections are "required" (i.e., removing them the algorithm failed and resulted in backtracking) or "non-required" (i.e., the constraints can be met without using these corrections) by means of two binary vectors, $\mathbf{r} = \{r_n;\, n = 1, ..., N\}$ and $\mathbf{s} = \{s_n;\, n = 1, ..., N\}$, whose $n$-th entry is $1$ if the $n$-th correction is marked as required/non-required and $0$, otherwise. Accordingly, the *CP* nature of the proposed methodology is not related to the compression of a matrix through singular value decomposition, but rather to the retrieval of the *sparsest* set of failure corrections $\Delta\mathbf{w}_{opt}$ to comply with the radiation constraints at hand. From an algorithmic viewpoint, the procedural steps of the *CP* method look as follows:

- *Step 0* [*Initialization*] ($k = 0$) - Reset the backtrack vectors ($\mathbf{r}^{(k)}\big\rfloor_{k=0} = \mathbf{0}$ and $\mathbf{s}^{(k)}\big\rfloor_{k=0} = \mathbf{0}$) and compute the initial solution, $\Delta\mathbf{w}_{opt}^{(k)}\Big\rfloor_{k=0}$ ($\Delta\mathbf{w}_{opt}^{(k)}$ being the current best solution of the *CP* algorithm at the $k$-th iteration), as the $\ell_1$-norm solution of the failure correction problem

$$\Delta\mathbf{w}_{opt}^{(k)}\Big\rfloor_{k=0} = \arg\min_{\Delta\mathbf{w}} \left\{ \left\|\Delta\mathbf{w}\right\|_1 \middle| \Phi\left(\Delta\mathbf{w}\right) \leq \Phi_{target} \text{ and } \Delta w_n \Omega_n = 0 \right\} \tag{7}$$

where $\left\|\cdot\right\|_1$ denotes the $\ell_1$-norm. Towards this end, the constrained minimization problem defined in (7) is solved with the interior-point algorithm [42]. This latter technique iteratively defines and solves (via a gradient descent search) a sequence of intermediate equality-constrained optimization problems which approximate the original one with increasing accuracy until convergence is reached [42]. More specifically, the procedure is initialized at $\Delta\mathbf{w} = \mathbf{0}$ and the iterations ($i$ being the interior-point algorithm iteration index, $i = 1, ..., I$) are stopped when at least one of the following conditions holds true: ($a$) the number of iterations exceeds $I_{max}$ ($i > I_{max}$), ($b$) a step smaller than $\xi$ is attempted, or ($c$) the first-order optimality condition is satisfied within the threshold value $\zeta$, where $I_{max}$, $\xi$, and $\zeta$ are user-defined control parameters.
- *Step 1* [*Least-Important Correction Guess*] - Increment the iteration index ($k \rightarrow k+1$) and find the position, $n_{least}$, of the correction within the last best solution, $\Delta\mathbf{w}_{opt}^{(k-1)}$, that is of "least importance". In the following, this choice is carried out by identifying the correction having the minimal non-zero magnitude (since, according to Parseval theorem, it is expected to have the minimum integral impact on the radiation pattern) and not marked as "required"

$$n_{least} = \arg\min_{n} \left\{ \left|\Delta w_{opt,n}^{(k-1)}\right| \middle| \Delta w_{opt,n}^{(k-1)} \neq 0 \text{ and } r_n^{(k-1)} \neq 0 \right\}. \tag{8}$$

If no such correction is found, then all corrections are either $0$ or marked as "required" and the iteration stops ($k = k_{opt}$) by setting the optimal set of excitation corrections to $\Delta\mathbf{w}_{opt} \equiv \Delta\mathbf{w}_{opt}^{(k-1)}$. Otherwise, the vector of "non-required" excitations is updated by adding the $n_{least}$-th correction ($s_n^{(k)} = s_n^{(k-1)} + \delta_{n,n_{least}}$, $\delta_{p,q}$ being the Kronecker delta) and a trial solution, $\Delta\mathbf{w}_{tr}^{(k)}$, is generated by removing the $n_{least}$-th correction from the last best solution

$$\Delta w_{tr,n}^{(k)} = \Delta w_{opt,n}^{(k-1)} \left(1 - \delta_{n,n_{least}}\right). \qquad (9)$$

It is worth noting that the trial solution $\Delta\mathbf{w}_{tr}^{(k)}$ has a $\ell_0$-norm value smaller than $\Delta\mathbf{w}_{opt}^{(k-1)}$ since $\Delta w_{tr,n}^{(k)} = 0$ if $n = n_{least}$, while $\Delta w_{tr,n}^{(k)} = \Delta w_{opt,n}^{(k-1)}$ otherwise, but the pattern constraints could be not fit since the array weights have been changed and, in turn, the pattern, as well;

- *Step 2* [*Correction Removal Attempt*] - By using $\Delta\mathbf{w}_{tr}^{(k)}$ as reference configuration, look for a new set of corrections, $\Delta\mathbf{w}_{opt}^{(k)}$, with minimal $\ell_1$-norm that meets the pattern requirements without changing any of the corrections that are faulty or marked as "non-required". Towards this end, the following constrained minimization problem

$$\Delta\mathbf{w}_{opt}^{(k)} = \arg\min_{\Delta\mathbf{w}} \left\{ \left\|\Delta\mathbf{w}\right\|_1 \middle| \Phi\left(\Delta\mathbf{w}\right) \leq \Phi_{target} \right.$$
$$\left. \text{and } \Delta w_n \left(\Omega_n + s_n^{(k)}\right) = 0 \right\} \qquad (10)$$

is solved still by means of the interior-point algorithm. If the attempt is successful, then the set of "required" excitations is cleared ($\mathbf{r}^{(k)} = \mathbf{0}$) and goto *Step 1*. Otherwise, goto *Step 3*;

- *Step 3* [*Backtrack*] - If the *Step 2* does not succeed in finding a new best vector, $\Delta\mathbf{w}_{opt}^{(k)}$, the *CP* algorithm backtracks on the last guess. This means that both the best correction vector and the vector of "non-required" corrections are kept from the previous iteration ($\Delta\mathbf{w}_{opt}^{(k)} \equiv \Delta\mathbf{w}_{opt}^{(k-1)}$, $\mathbf{s}^{(k)} \equiv \mathbf{s}^{(k-1)}$), while the $n_{least}$-th correction is marked as required ($r_n^{(k)} = r_n^{(k-1)} + \delta_{n,n_{least}}$). Goto *Step 1*.

It is worth noticing that the $\ell_0$-norm of the corrections is never minimized directly by the interior point algorithm, which rather operates on the $\ell_1$-norm expression in (10). Moreover, alternative concepts may be used to implement the heuristic for the selection of the "least important correction" in Step. 1. Indeed, a universal heuristic cannot be *a-priori* defined since the relevance of a correction strongly depends on the user-defined $\Phi(\cdot)$. Nevertheless, according to the Parseval theorem, the proposed heuristic (also adopted in [37] in a different context) is expected to provide robust performance whenever $\Phi(\cdot)$ measures a feature depending on the array pattern shape (such as *SLL* or *HPBW*).

In order to detail the step-by-step behaviour of the *CP* method, let us consider a "toy" example where a simplified low-dimension correction problem is dealt with. More in detail, the scenario at hand is that of an array with $N = 4$ elements subject to a failure on the $n = 2$ element (i.e., $N_F = 1$ and $\Omega_2 = 1 \rightarrow N_C = 3$), where the original set of excitations has been chosen so that the radiated pattern is a Dolph-Chebyshev (*DC*) one with a *SLL* value equal to $SLL = -15$ [dB] (i.e., $\mathbf{w} = \{1, 0.419, 0.419, 1\}$). Moreover, the metric of the array performance, $\Phi$, has been set to the maximum of the *SLL* [$SLL(u) \triangleq 10 \times \log_{10}\left(\frac{\widetilde{PP}(u)}{\widetilde{PP}(0)}\right)$, being $PP(u) \triangleq |F(u)|^2$]

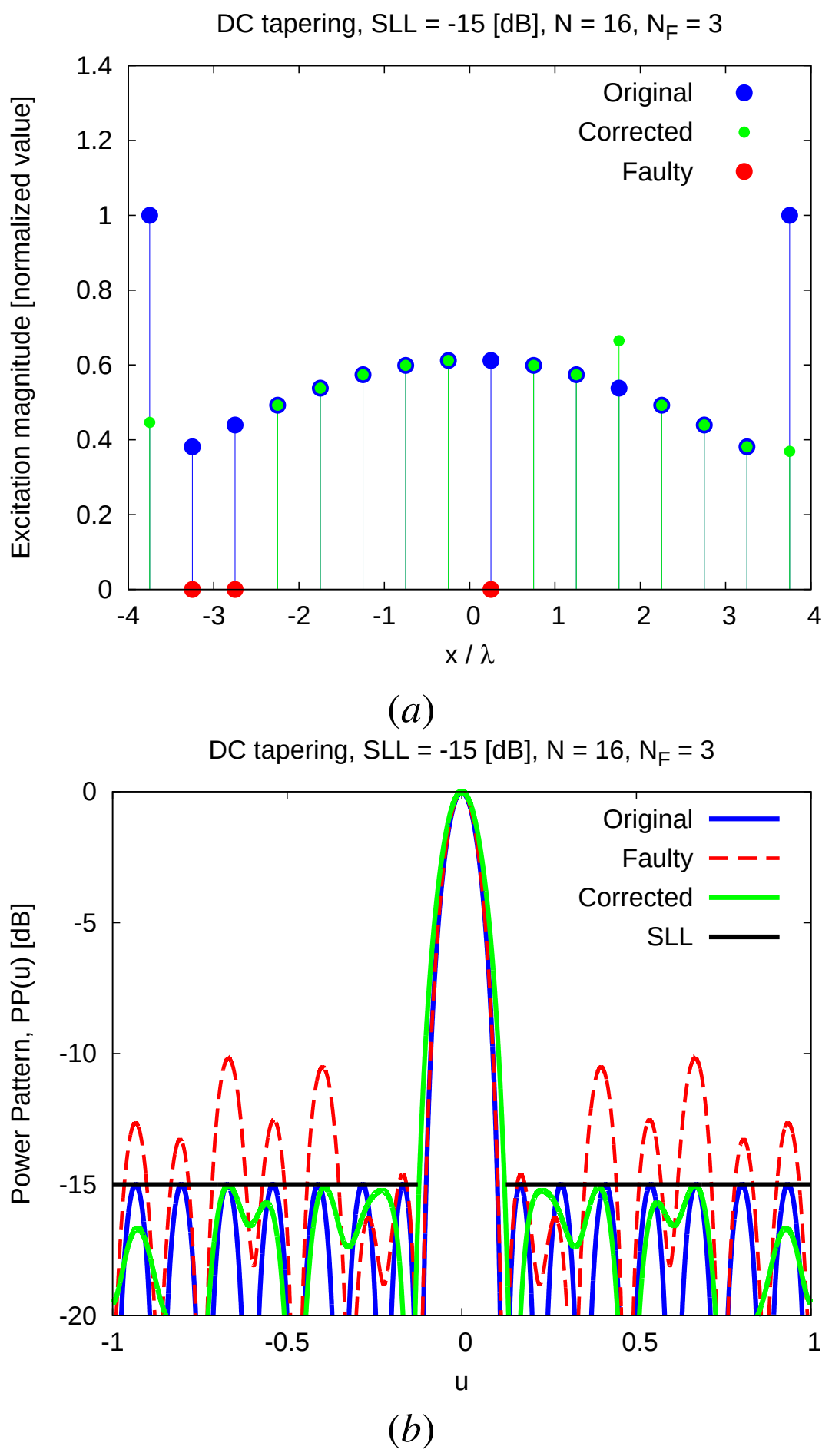

Figure 3. *Numerical Assessment* (*Test Case #1*: $N = 16$, $N_F = 3 \rightarrow \eta_F \approx 19$ %; *DC SLL* $= -15$ [dB]) - Plot of (*a*) the excitations and of (*b*) the power patterns of the original, the faulty, and the corrected arrays.

within the angular set $\mathbf{\Theta}$ ($\mathbf{\Theta} = \{u = \pm 0.5,\ u = \pm 0.7\}$)

$$\Phi\left(\Delta\mathbf{w}\right) = \max_{u \in \mathbf{\Theta}} \left\{SLL\left(u\right)\right\} \qquad (11)$$

with a target value equal to $\Phi_{target} \equiv SLL_{target} = -5.5$ [dB].

Figure 2 and Table I summarize the application of the *CP* algorithm to the "toy" problem at hand. At the initialization ($k = 0$), the solution $\Delta\mathbf{w}_{opt}^{(0)} = \{-0.438,\ 0.0,\ 0.593,\ -9.72 \times 10^{-6}\}$ [Tab. I - Fig. 2(*a*)] has been found by performing the $\ell_1$-norm minimization (7). Successively ($k = 1$), first the correction at the $n = 4$ array position has been identified as the least important ($n_{least} = 4$), since it has the smallest non-zero magnitude within $\Delta\mathbf{w}_{opt}^{(k-1)}$. Thus, a trial solution has been generated by re-setting the $n_{least}$-th correction of $\Delta\mathbf{w}_{opt}^{(0)}$ ($\Delta\mathbf{w}_{tr}^{(1)} = \{-0.438,\ 0.0,\ 0.593,\ 0.0\}$) and updating the "non required" vector $\mathbf{s}^{(1)}$ [$s_4^{(1)} = 1$ - Fig. 2(*b*)] (*Step 1*). Afterward (*Step 2*), since $\Delta\mathbf{w}_{tr}^{(1)}$ already satisfies the pattern requirements [$\Phi(\Delta\mathbf{w}_{tr}^{(1)}) \leq \Phi_{target}$], the $\ell_1$-norm minimization (7) has not been necessary and the current best solution has been updated to the trial one [$\Delta\mathbf{w}_{opt}^{(k)} \equiv \Delta\mathbf{w}_{tr}^{(1)}$ - Fig. 2(*a*)] by continuing the iterative process with the *Step 1*, once again. At the next iteration ($k = 2$), the correction at $n = 1$ has been guessed to be the least important one ($n_{least} = 1$) and $\mathbf{s}^{(2)}$ has been updated by setting $s_1^{(2)} = 1$ [Fig. 2(*b*)]. Moreover, a trial solution has been generated accordingly ($\Delta\mathbf{w}_{tr}^{(2)} = \{0.0, 0.0, 0.593, 0.0\}$ - Tab. I) to start the $\ell_1$-norm minimization of the *Step 2*. A new solution has been generated, $\Delta\mathbf{w}^{(2)} = \{-0.438, 0.0, 0.593, 0.0\}$, that fits the pattern requirements [$\Phi(\Delta\mathbf{w}^{(2)}) \leq \Phi_{target}$], then the current best solution has been updated [$\Delta\mathbf{w}_{opt}^{(2)} \equiv \Delta\mathbf{w}^{(2)}$ - Fig. 2(*a*)]. When running the iteration $k = 3$, since $s_1^{(3)} = s_4^{(3)} = 1$ and $\Omega_2 = 1$ [Fig. 2(*b*)], the location of the least important correction has been mandatorily set to $n_{least} = 3$. However, removing the correction at $n = 3$ in $\Delta\mathbf{w}_{opt}^{(2)}$ meant no correction to be done (i.e., $\Delta\mathbf{w}_{tr}^{(3)} = \{0.0, 0.0, 0.0, 0.0\}$), thus the generation of an unreliable pattern [$\Phi(\Delta\mathbf{w}_{tr}^{(3)}) > \Phi_{target}$]. Therefore, the algorithm has backtracked on its decision by removing the correction on the $n = 3$ element of the array from the non-required corrections ($s_3^{(3)} = 0$) and marking it as required [$r_3^{(3)} = 1$ - Fig. 2(*b*)]. The algorithm reached the last iteration ($k = 4$), but halted at the *Step 1* because all corrections are either zero or marked as "required" ($\rightarrow \Delta\mathbf{w}_{opt} \equiv \Delta\mathbf{w}_{opt}^{(3)}$ and $\widehat{N_C} = 1$).

## IV. Numerical Assessment and Validation

This Section describes a representative set of numerical benchmarks that have been performed to assess the effectiveness as well as the limitations of the *CP*-based correction also with respect to reference methods in the state-of-the-art literature [21][23].

For comparison purposes, but without loss of generality, the function $\Phi$ has been set to the *SLL* of the radiated pattern (11) in the sidelobe angular region $\boldsymbol{\Theta}\,(BW_{target})$, $BW_{target}$ being - unless mentioned otherwise - the mainlobe beamwidth of the pattern radiated by an array with $N = N_C$ elements. More specifically, the mainlobe beamwidth, $BW$, is defined as the angular range for which $PP\,(u) \geq SLL_{target}$. Finally, the target threshold $\Phi_{target}$ has been generally chosen equal to the *SLL* of the original whole array ($\Phi_{target} \equiv SLL_{target} = SLL$).

### A. Numerical Assessment

The first test case (*Test Case #1* - Tab. II) deals with an array whose dimension allows one the computation of the optimal solution to the *MCFC* problem (6) through an exhaustive process[1] in a limited amount of time. It is concerned with a $N = 16$ elements Dolph-Chebyshev (*DC*) tapered uniform linear array that radiates a pattern with an *SLL* equal to $SLL = -15$ [dB]. The failures have been assumed at $N_F = 3$ ($\rightarrow \eta_F = 18.75$ %, $\eta_F \triangleq \frac{N_F}{N}$) element locations ($\Omega_2 = \Omega_3 = \Omega_9 = 1$) [Fig. 3(*a*)] so that the *SLL* of the faulty array increases up to $\widetilde{SLL} = -10.19$ [dB] [Fig. 3(*b*)].

By comparing the results from the exhaustive search and the *CP* method (Fig. 4), it turns out that they are coincident with $\widehat{N_C} = 3$ ($\rightarrow \eta_C \equiv \eta_F$, $\eta_C \triangleq \frac{N_C}{N}$) excitations being corrected [Fig. 3(*a*) and Fig. 4(*a*)] and a radiation pattern having the same (rather smaller) *SLL* of the original array [Fig. 3(*b*) and Tab. III], while maintaining the beamwidth within $\text{BW}_{target} = 14.6$ [deg] (i.e., the beamwidth of a *DC*-tapered uniform linear array of $N = N_C$ elements), despite the use of only $\widehat{\eta}'_C = 23.08$ % ($\widehat{\eta}'_C \triangleq \frac{\widehat{N_C}}{N_C}$) of the admissible corrections, $N_C$. Table III summarizes the outcomes by also reporting the values of both the $\ell_0$- ($\equiv \widehat{N_C}$) and the $\ell_1$-norm of the correction vector.

[1] It means that the constrained minimization problem in (7) is solved for each value of $\widehat{N_C}$ ($\widehat{N_C} = 1, ..., N_C$), that is $\sum_{\widehat{N_C}=1}^{N_C} \binom{N_C}{\widehat{N_C}}$ times.

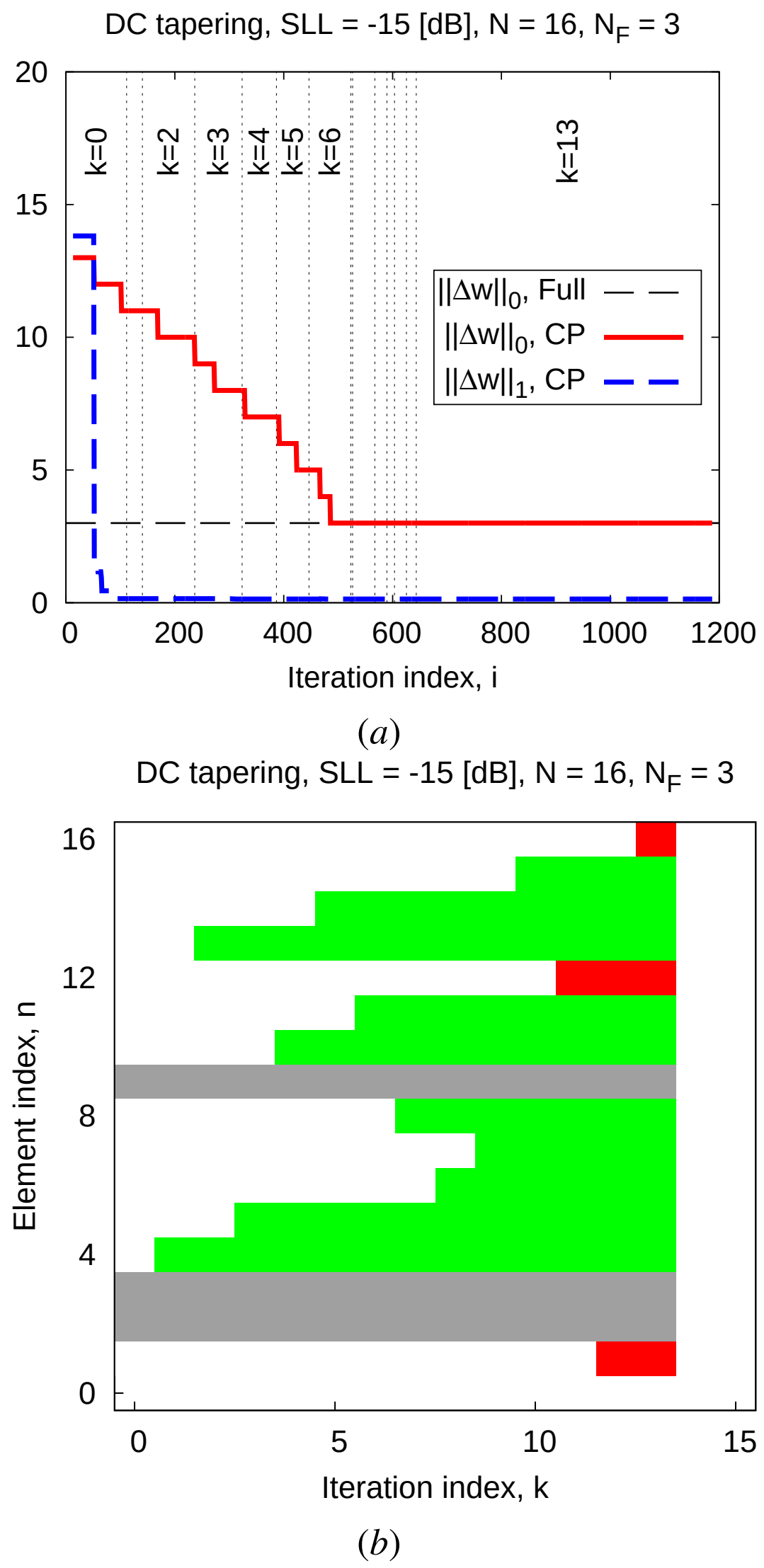

Figure 4. *Numerical Assessment* (*Test Case #1*: $N = 16$, $N_F = 3 \rightarrow \eta_F \approx 19$ %; *DC SLL* $= -15$ [dB]) - Evolution of (*a*) the $\ell_1$-norm and the $\ell_0$-norm of the excitation correction vector, $\Delta\mathbf{w}$, and of (*b*) the status ("required/non-required") of the element corrections, $r_n/s_n$ ($n = 1, ..., N$), versus the *CP* iteration index, $k$.

Table II
*Numerical & Comparative Assessment* - Test cases descriptors.

| *Test Case* | $N$ | *Faulty Element Indexes* | $\eta_F$ |
|---|---|---|---|
| 1 | 16 | $\{2, 3, 9\}$ | 18.75 |
| 2 | 50 | $\{8, 13, 38\}$ | 6.0 |
| 3 | 32 | $\{2, 5\}$ | 6.25 |
| 4 | 32 | $\{2, 5, 6\}$ | 9.375 |

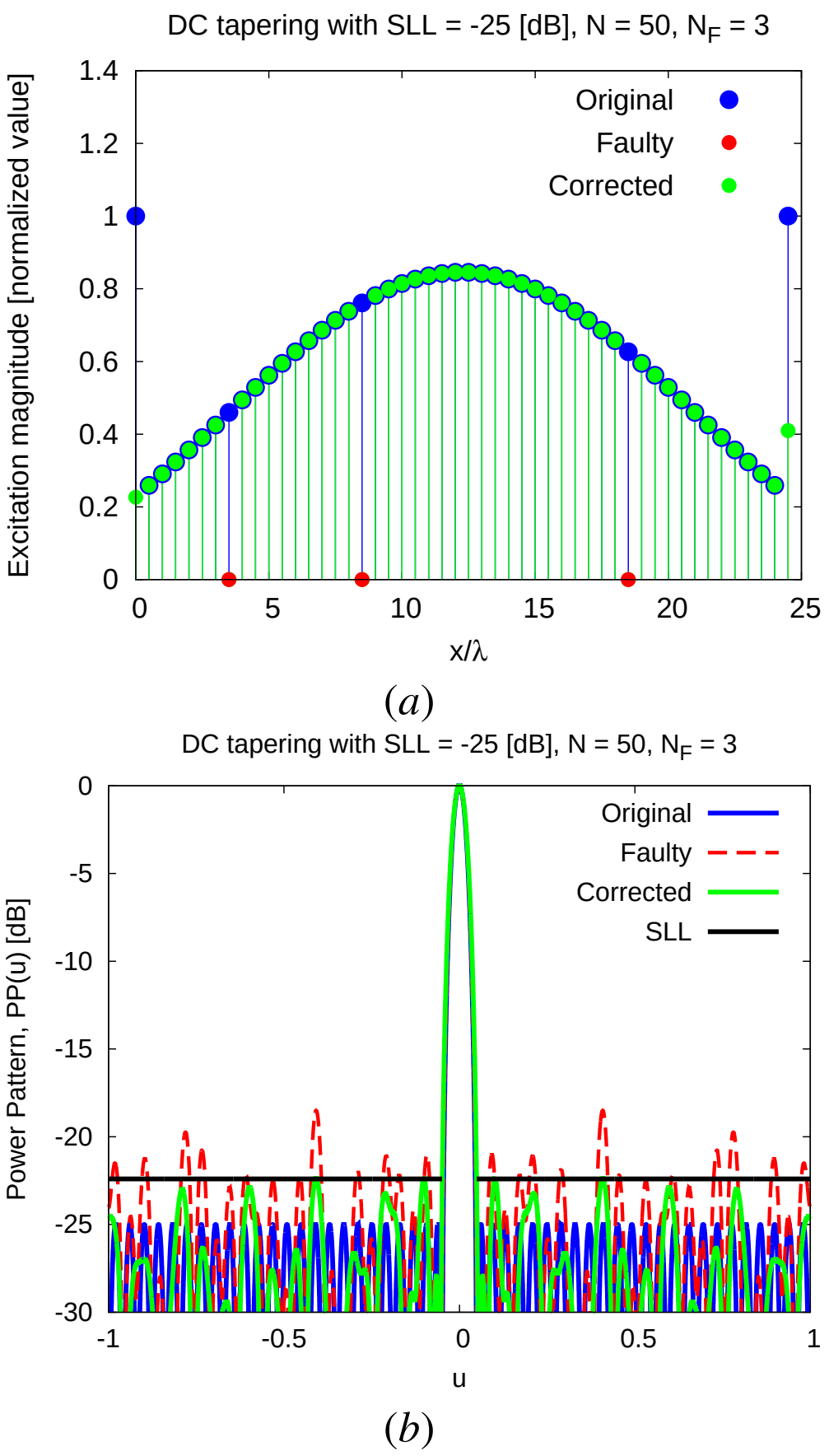

Figure 5. *Comparative Assessment* (*Test Case #2*: $N = 50$, $N_F = 3 \rightarrow \eta_F = 6$ %; *DC SLL* $= -25$ [dB]; $SLL_{target} = -22.4$ [dB]) - Plot of ($a$) the excitations and of ($b$) the power patterns of the original, the faulty, and corrected arrays either in [23] or synthesized with the *CP* method.

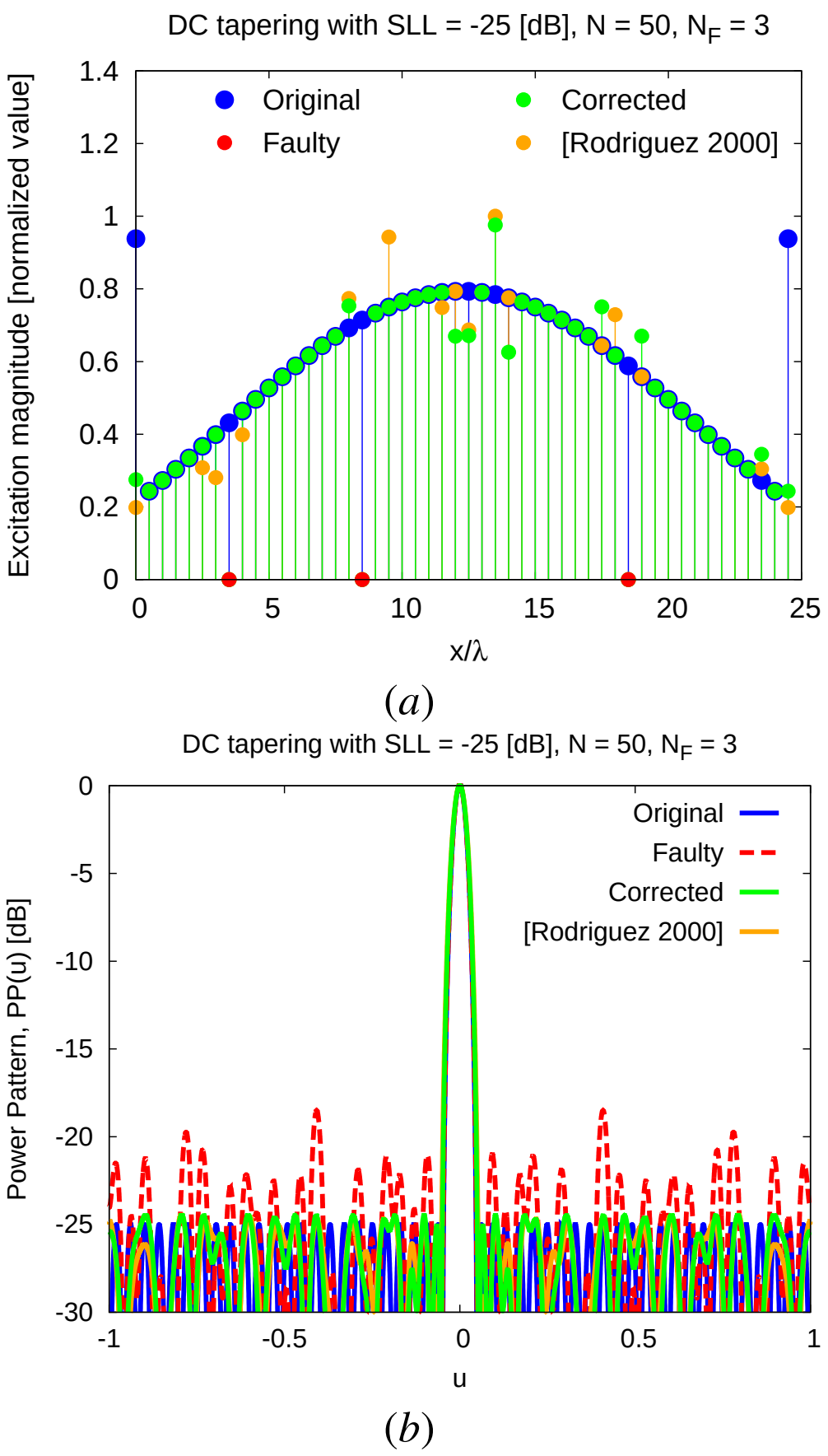

Figure 6. *Comparative Assessment* (*Test Case #2*: $N = 50$, $N_F = 3 \rightarrow \eta_F = 6$ %; *DC SLL* $= -25$ [dB]; $SLL_{target} = -24.5$ [dB]) - Plot of ($a$) the excitations and of ($b$) the power patterns of the original, the faulty, and corrected arrays either in [23] or synthesized with the *CP* method.

Table III
*Numerical Assessment* (*Test Case #1*: $N = 16$, $N_F = 3 \rightarrow \eta_F \approx 19$ %; *DC SLL* $= -15$ [dB]) - PERFORMANCE INDEXES.

| | ***SLL*** **[dB]** | ***BW*** **[deg]** | $\lVert \Delta \mathbf{w} \rVert_1$ | $\lVert \Delta \mathbf{w} \rVert_0$ |
|---|---|---|---|---|
| *Target* | $-15.0$ | 14.6 | − | − |
| *Original Array* | $-15.00$ | 11.70 | − | − |
| *Faulty Array* | $-10.19$ | 11.83 | − | − |
| *Corrected Array* | $-15.02$ | 14.15 | 1.31 | 3 |

To analyze the behavior of the *CP* correction method more in detail, Figure 4 shows the evolution of the norms of the current best solution, $\mathbf{w}_{opt}^{(k)}$ [Fig. 4($a$)], as well as of the backtrack vectors [Fig. 4($b$)] versus the iteration index $k$. According to the *CP* guidelines and starting from the initial setup ($\mathbf{r}^{(k)}\rfloor_{k=0} = \mathbf{0}$ and $\mathbf{s}^{(k)}\rfloor_{k=0} = \mathbf{0}$), the iterative process marks each array element as "required" or "non-required" until all elements, unless the faulty ones, are labeled [Fig. 4($b$)]. Unless the beginning, each time a correction turns out to unnecessary ($s_n^{(k)} = 1$), the number of corrections, $\widehat{N_C}\Big\rfloor_k$ ($\widehat{N_C}\Big\rfloor_k \equiv \left\lVert \Delta \mathbf{w}_{opt}^{(k)} \right\rVert_0$), decreases by a unit, while $\left\lVert \Delta \mathbf{w}_{opt}^{(k)} \right\rVert_1$ seems unaffected since, when the $\ell_1$-norm minimization (7) takes place, most corrections already have a very small value ($|\Delta w_n| \approx 0$), thus removing them does not bring a visible variation.

## B. Comparative Assessment

As for the comparative study with competitive state-of-the-art methods, the "*Test Case #2*" refers to a benchmark example from [23]. More specifically, a $N = 50$ elements linear has been considered with $N_F = 3$ ($\eta_F = 6$ %) faulty elements as indicated in Tab. II and shown in Fig. 5($a$) where the array excitations are reported. Due to the failures, the *SLL* of the array worsens from $SLL = -25$ [dB] up to $\widetilde{SLL} = -18.51$ [dB] [Fig. 5($b$)]. When setting the target *SLL* to the same value yielded in [23] with a minimum of 5 corrections, it turns out that the *CP* solution recovers the target features by perturbing only $\widehat{\eta_C'} = 4.26$ % of the non-defective elements (vs. $\widehat{\eta_C'} = 10.64$ % in [23] - Tab. IV), the number of corrections ($\widehat{N_C} = 2$) being smaller than the number of faulty elements, as well. When reducing the threshold down to $SLL_{target} = -24.50$ [dB], the number of compensating elements increases to $\widehat{N_C} = 10$ [vs. $\widehat{N_C}^{[Rodriguez\ 2000]} = 12$ - Fig. 6($a$)] with a value of the amplitude distribution coefficient ($DR \triangleq max_{n=1,...,N}\left|\frac{I_n}{I_{n\pm 1}}\right|$) better than the original one ($DR^{CP} = 1.56$ vs. $DR = 3.86$) as for the *GA*-based correction ($DS^{[Rodriguez\ 2000]} = 1.29$) (Tab. IV). It is also worth pointing out that the *CPU*-time for the correction process takes less than 2 minutes (i.e., $\Delta t = 75$ [sec], while $\Delta t = 108$ [sec] for the case with $SLL_{target} = -22.40$ [dB]) on a standard laptop

Table IV
*Comparative Assessment* (*Test Case* #2: $N = 50$, $N_F = 3 \rightarrow \eta_F = 6$ %; *DC* $SLL = -25$ [dB]; $SLL_{target} = -24.5$ [dB]) - PERFORMANCE INDEXES.

| | ***SLL* [dB]** | ***BW* [deg]** | ***DR*** | $\|\Delta\mathbf{w}\|_1$ | $\|\Delta\mathbf{w}\|_0$ | $\eta_C$ [%] | $\widehat{\eta}'_C$ [%] |
|---|---|---|---|---|---|---|---|
| *Target* | $-22.4$ | 6.0 | – | – | – | – | – |
| *Original Array* | $-25$ | 5.18 | 3.86 | – | – | – | – |
| *Faulty Array* | $-18.51$ | 5.25 | 3.86 | – | – | 94.0 | 0.0 |
| *Corrected Array* | $-22.4$ | 5.60 | 1.58 | 1.36 | 2 | 94.0 | 4.26 |
| [Rodriguez 2000] | $-22.4$ | *n.a.* | *n.a.* | *n.a.* | 5 | 94.0 | 10.64 |
| *Target* | $-24.50$ | 5.66 | – | – | – | – | – |
| *Original Array* | $-25.00$ | 5.18 | 3.86 | – | – | – | – |
| *Faulty Array* | $-18.51$ | 5.25 | 3.86 | – | – | 94.0 | 0.0 |
| *Corrected Array* | $-24.50$ | 5.53 | 1.56 | 2.45 | 10 | 94.0 | 21.3 |
| [Rodriguez 2000] | $-24.47$ | 5.76 | 1.29 | 2.67 | 12 | 94.0 | 25.5 |

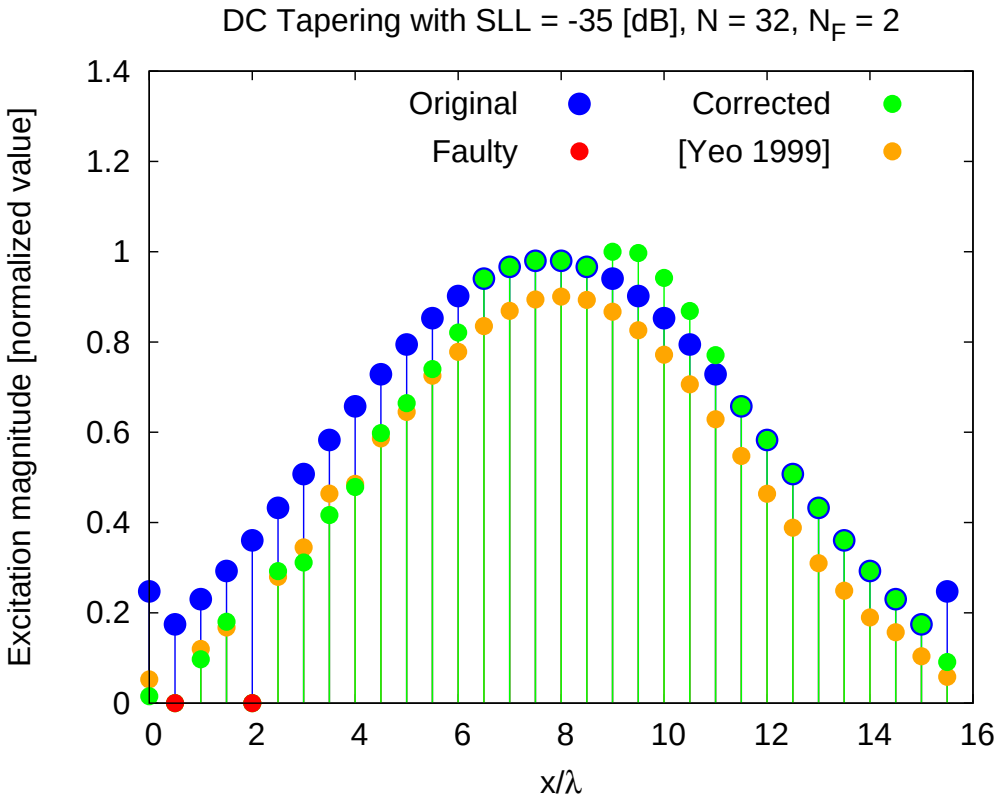

(*a*)

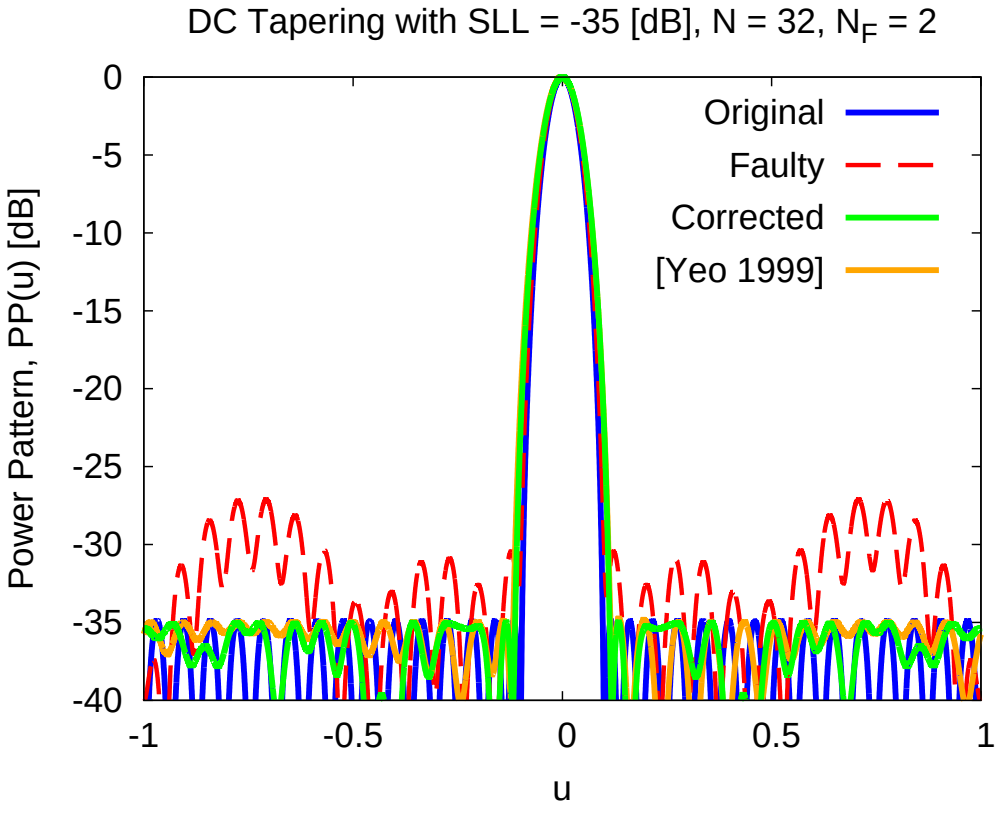

(*b*)

Figure 7. *Comparative Assessment* (*Test Case* #3: $N = 32$, $N_F = 2 \rightarrow \eta_F = 6.25$ %; *DC* $SLL = -35$ [dB]) - Plot of (*a*) the excitations and of (*b*) the power patterns of the original, the faulty, and corrected arrays either in [21] or synthesized with the *CP* method.

computer.

Still for comparison purposes, the *CP* method is applied to the "*Test Case #3*" (Tab. II) drawn from [21] where the authors proposed a *FC* method not aimed at reducing the number of corrections (i.e., $\widehat{N_C}^{[Yeo\,1999]} = N_C$), but devoted to yield the optimal results in terms of *SLL* and beamwidth for the corrected array. In [21], the original array was a $N = 32$-elements linear array with a Dolph-Chebyshev tapering and $SLL = -35$ [dB], while $N_F = 2$ ($\rightarrow \eta_F = 6.25$ %) failures have been applied ($\Omega_2 = \Omega_5 = 1$) causing a performance downgrade of about 8 [dB] in the *SLL* (i.e., $\widetilde{SLL} = -27.11$ [dB]).

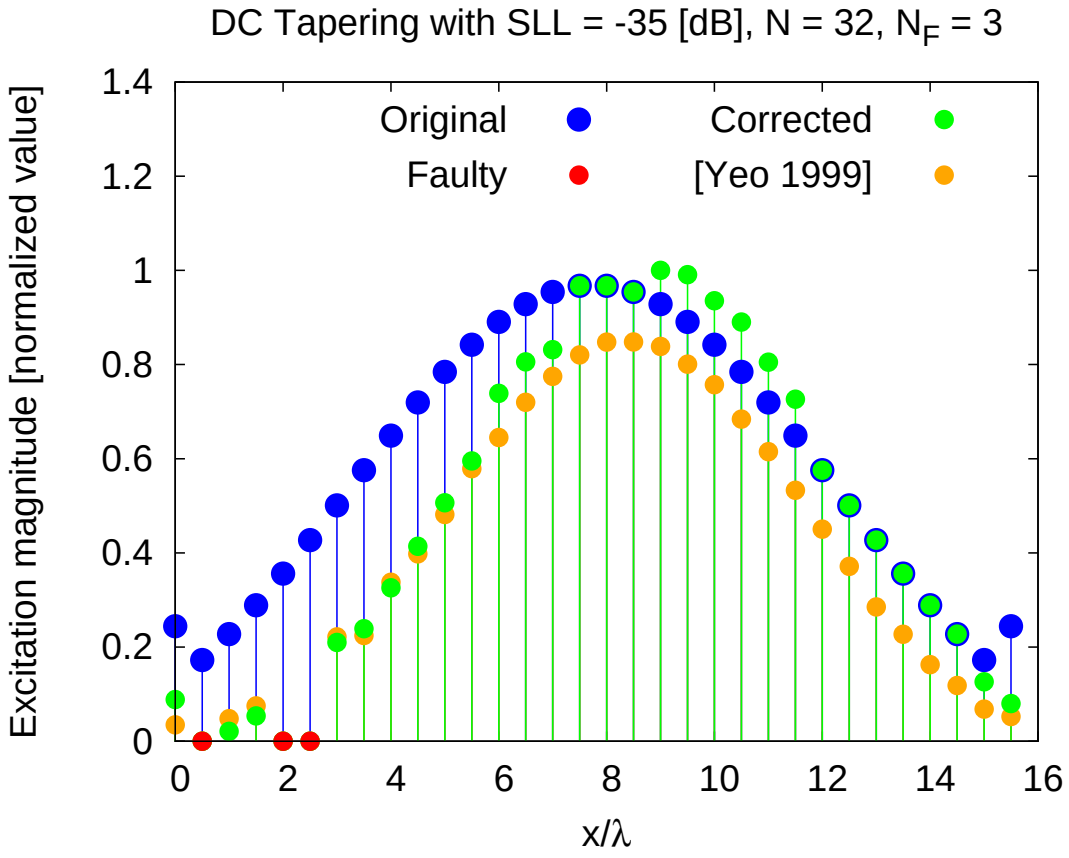

(*a*)

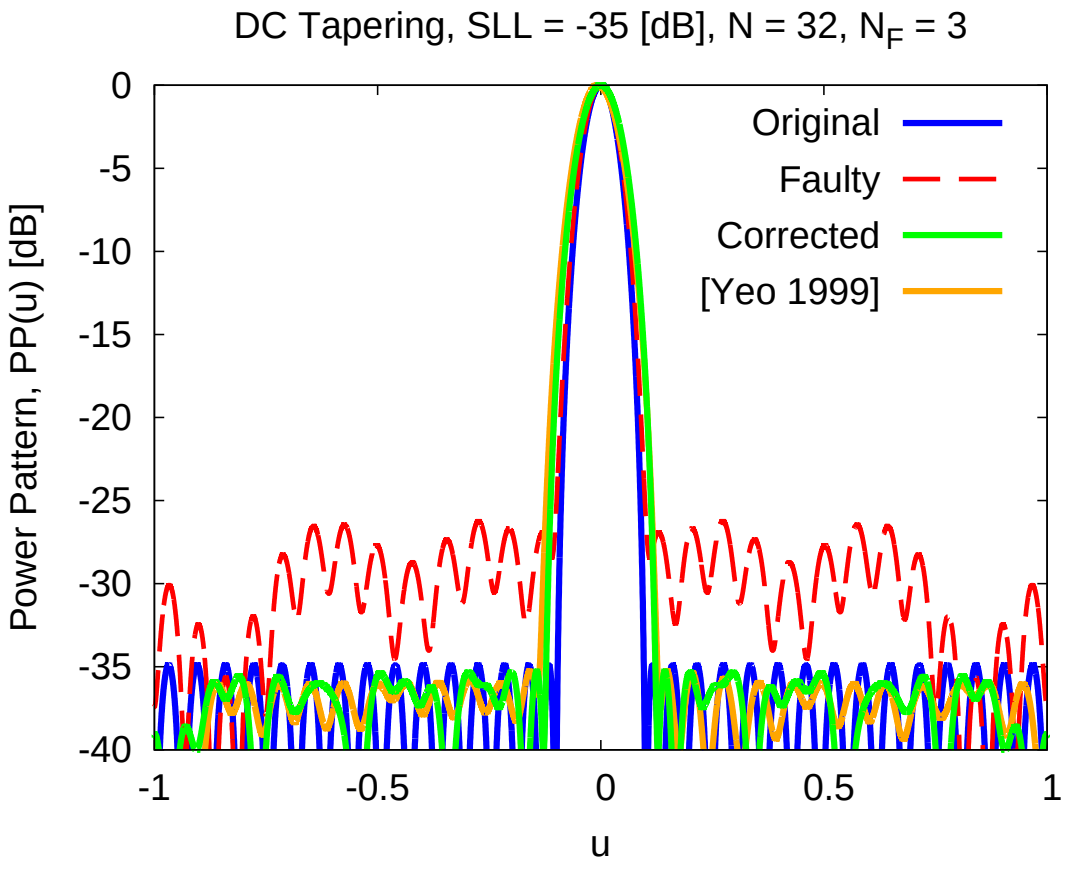

(*b*)

Figure 8. *Comparative Assessment* (*Test Case* #4: $N = 32$, $N_F = 3 \rightarrow \eta_F \approx 9.4$ %; *DC* $SLL = -35$ [dB]; $SLL_{target} = -35.28$ [dB]) - Plot of (*a*) the excitations and of (*b*) the power patterns of the original, the faulty, and corrected arrays either in [21] or synthesized with the *CP* method.

Table V
*Comparative Assessment* (*Test Case* #3: $N = 32$, $N_F = 2 \rightarrow \eta_F = 6.25$ %; *DC* $SLL = -35$ [dB]; *Test Case #4*: $SLL_{target} = -35.28$ [dB]) - PERFORMANCE INDEXES.

| | ***SLL* [dB]** | ***HPBW* [deg]** | $\|\Delta\mathbf{w}\|_1$ | $\|\Delta\mathbf{w}\|_0$ | $\eta_C$ [%] | $\widehat{\eta}'_C$ [%] |
|---|---|---|---|---|---|---|
| *Target* | $-35.00$ | – | – | – | – | – |
| *Original Array* | $-35.00$ | 4.15 | – | – | – | – |
| *Faulty Array* | $-27.11$ | 4.29 | – | – | 93.75 | 0.0 |
| *Corrected Array* | $-35.00$ | 4.74 | 2.17 | 17 | 93.75 | 56.66 |
| [Yeo 1999] | $-34.78$ | 4.77 | 3.57 | 30 | 93.75 | 100.0 |
| *Target* | $-35.28$ | – | – | – | – | – |
| *Original Array* | $-35.00$ | 4.15 | – | – | – | – |
| *Faulty Array* | $-26.24$ | 4.38 | – | – | 90.625 | 0.0 |
| *Corrected Array* | $-35.28$ | 5.34 | 3.86 | 20 | 90.625 | 68.97 |
| [Yeo 1999] | $-35.28$ | 5.36 | 5.62 | 29 | 90.625 | 100.0 |

Figure 7(*a*) shows the original and the faulty excitations of the array along with the corrected ones reported in [21] or synthesized with the *CP* method. As it can be noted, this latter uses only $\widehat{N}_C = 17$ out of the $N_C = 30$ working elements ($\rightarrow \widehat{\eta}'_C = 56.66$ % - Tab. V) to fit the target requirements ($SLL_{target} = -35$ [dB]) by also slightly improving both the half-power beamwidth and the *SLL* of the corrected array in [21] [$HPBW^{CP} = 4.74$ [deg] vs. $HPBW^{[Yeo\,1999]} = 4.77$ [deg] and $SLL^{CP} = -35.0$ [dB] vs. $SLL^{[Yeo\,1999]} = -34.79$ [dB] - Tab. V and Fig. 7(*b*)].

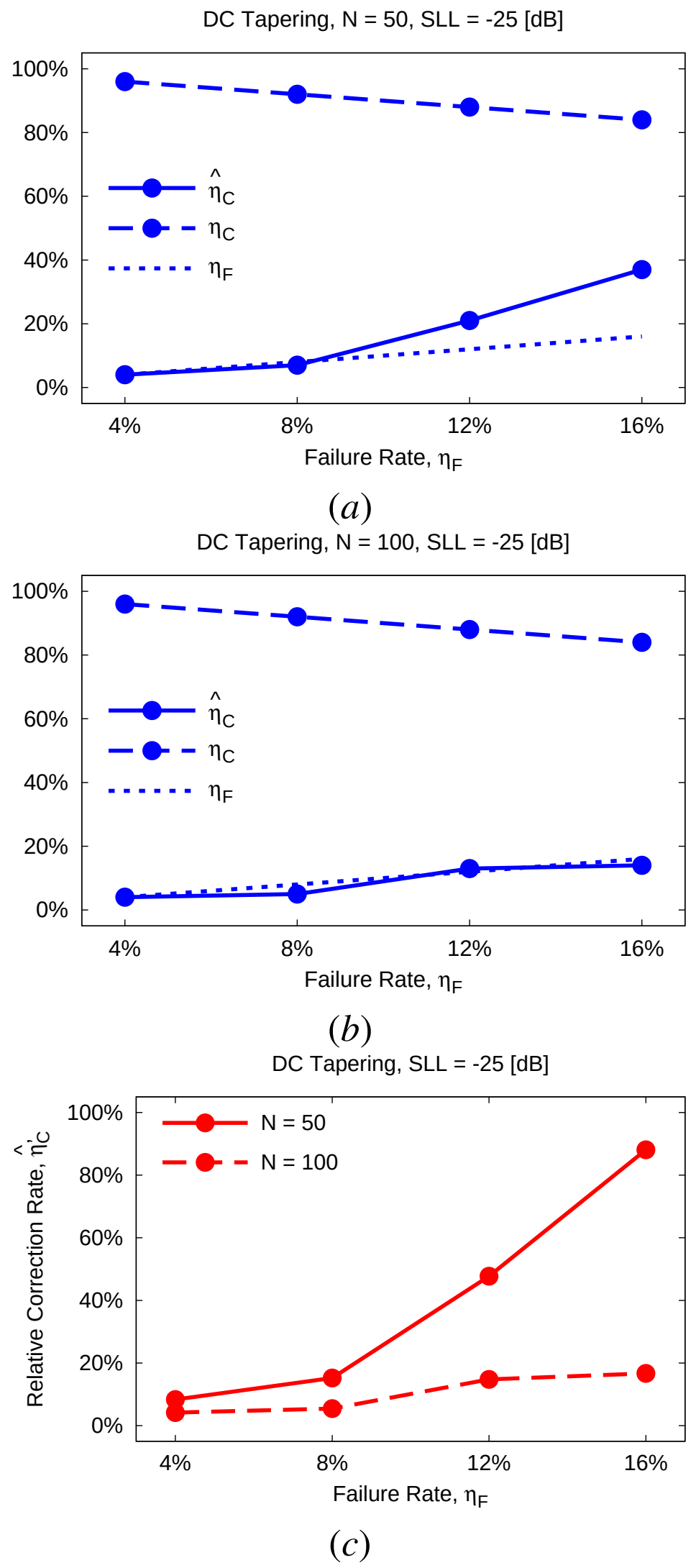

Figure 9. *Performance Analysis* (*DC* $SLL = -25$ [dB]) - Plots of the percentage of failed, $\eta_F$, reconfigurable, $\eta_C$, and corrected, $\widehat{\eta}_C$, elements versus the failure rate, $\eta_F$ in (*a*) a medium ($N_1 = 50$) and in (*b*) a large ($N_2 = 100$) array; (*c*) Behavior of the relative correction rate, $\widehat{\eta}'_C$, versus $\eta_F$.

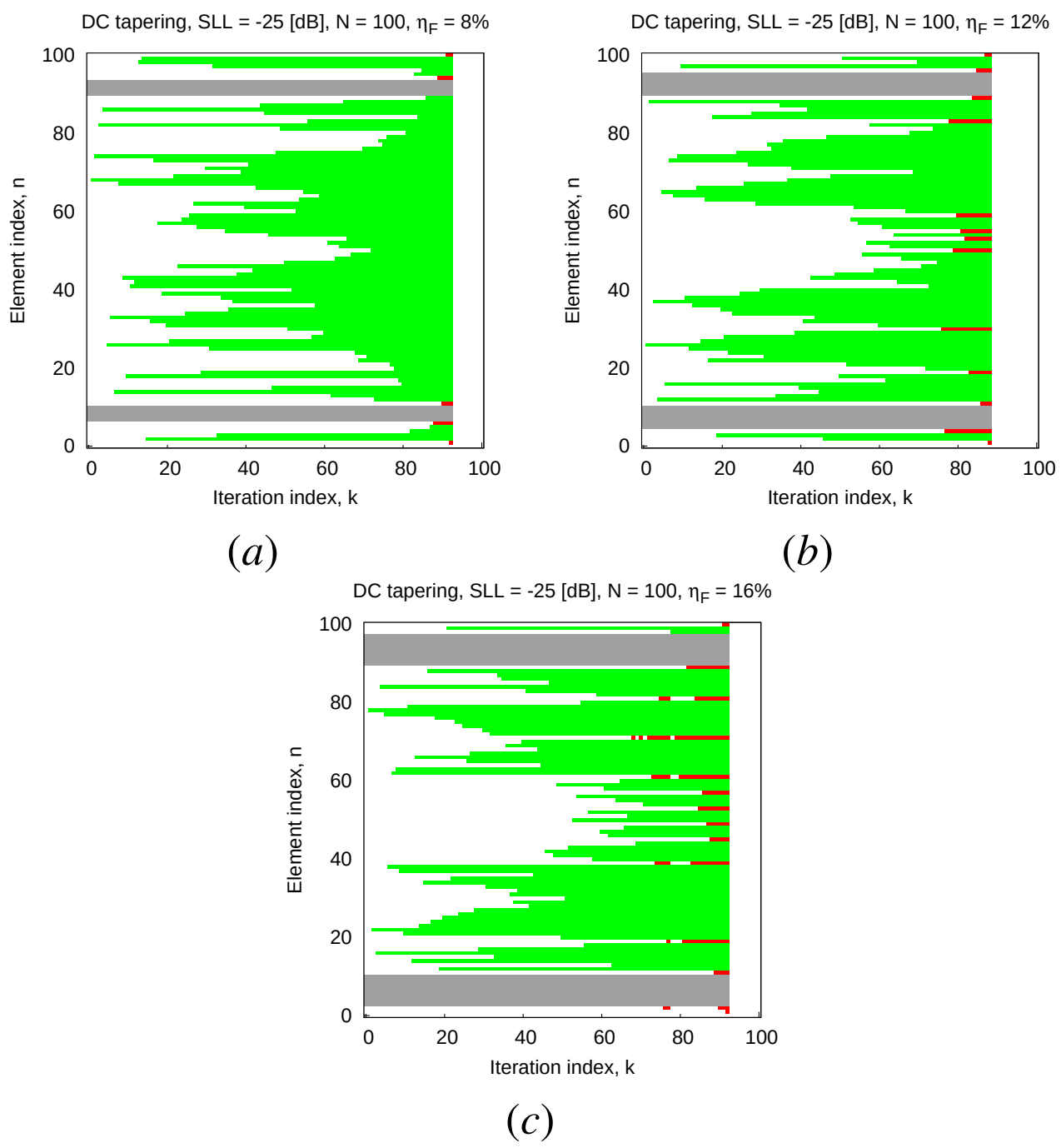

Figure 10. *Performance Analysis* ($N_2 = 100$; *DC* $SLL = -25$ [dB]) - Evolution of the backtracking vectors versus the *CP* iteration index when (*a*) $\eta_F = 8$ %, (*b*) $\eta_F = 12$ %, and (*c*) $\eta_F = 16$ %.

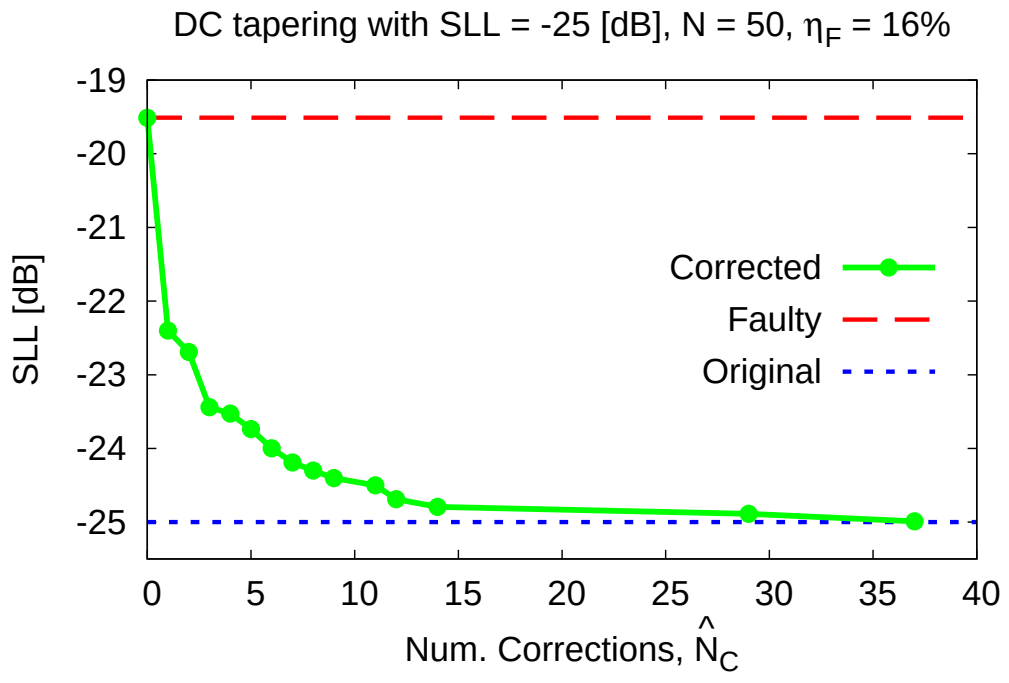

Figure 11. *Performance Analysis* (*DC* $SLL = -25$ [dB]; $N = 50$; $\eta_F = 16\%$) - Behavior of the side lobe level of the corrected array, $\widehat{SLL}$, versus the number of corrections applied, $\widehat{N}_C$.

In the next test case ("*Test Case #4*" - Tab. II), a failure has been added to the previous array configuration and it still comes from [21]. More specifically, the failure rate has been increased up to $\eta_F \approx 9.4$ % with another faulty element at $x = x_6$ (i.e., $\Omega_6 = 1$) that causes a further worsening of the *SLL* to $\widehat{SLL} = -26.24$ [dB] with respect to the whole array ($SLL = -35$ [dB]). For a fair comparison, since the correction method in [21] yielded an *SLL* equal to $SLL^{[Yeo\,1999]} = -35.28$ [dB], the same value has been maintained as target for the *CP*-based correction ($SLL_{target} \equiv SLL^{[Yeo\,1999]}$).

Once again, the *CP* method was successful in fitting the requirement ($SLL^{CP} = SLL_{target}$ - Tab. V) with a reduced number of elements [$\widehat{N}_C = 20$ - Fig. 8(*a*) and Tab. V], even though higher than the "*Test Case #3*" ($\widehat{\eta}'_C\rfloor_{N_F=3} > \widehat{\eta}'_C\rfloor_{N_F=2}$ - Tab. V) because of the presence of one more fault element. Moreover, the *CP* array affords a narrower beam than that of the corrected array in [21] [$HPBW^{CP} < HPBW^{[Yeo\,1999]}$ - Fig. 8(*b*) and Tab. V]. As far as the computational burden is concerned, the corrections for both cases (i.e., "*Test Case #3*" and "*Test Case #4*") have been synthesized by the *CP*-based approach in around 1 [min] (i.e., $\Delta t^{Test\,\#3} = 47$ [sec] and $\Delta t^{Test\,\#4} = 45$ [sec]).

## C. Performance Analysis

In order to assess the potentialities and the limitations of the proposed approach, a performance analysis has been carried out by varying the "correction scenario" at hand, but keeping always the array size from medium to large.

The first test case of this section deals with the dependence of the *CP* performance on the failure rate $\eta_F$. More in detail, two arrays ($N_1 = N$ and $N_2 = M \times N$, being $N = 50$ and $M = 2$) with DC tapering and $SLL = -25$ [dB] have been considered, while the failure rate has been varied within the range $\eta_F \in$ $\{p \times \eta_0;\, p = 1, ..., P\}$ ($\eta_0 = 2$ % and $P = 3$). The descriptive parameters of the benchmarks at hand are reported in Tab. VI along with the indexes of the correction outcomes. It is worth pointing out that the locations of the faulty elements (Tab. VI) have been chosen so that $\widetilde{SLL}\Big\rfloor_{N_1}^{\eta_F} \approx \widetilde{SLL}\Big\rfloor_{N_2}^{\eta_F}$ (i.e., elements of the same "importance" turn out to be switched off in both $N_1$- and $N_2$-elements arrays for a fixed value of $\eta_F$). Towards this end, all faulty arrangements share the condition $\Omega_5 = \Omega_{45} = 1$. Moreover, the following "*failure rule*" has been applied symmetrically to the array center, $n_f$ being the failure index ($f = 1, ..., F$). If $n_f < \frac{N}{2}$ then $n_{f-j}\rfloor_{M \times N} = M \times n_f\rfloor_N - j$, else ($n_f > \frac{N}{2}$) $n_{f-j}\rfloor_{M \times N} = M \times n_f\rfloor_N + j$ where $j = 0, ..., P \times N_F$.

Table VI
*Performance Analysis* (*DC* $SLL = -25$ [dB]) - Test cases descriptors and performance indexes.

| $N$ | *Faulty Element Indexes* | $SLL$ | $\widetilde{SLL}$ | $SLL_{target}$ | $SLL^{CP}$ | $BW$ | $\widetilde{BW}$ | $BW_{target}$ | $BW^{CP}$ | $\widehat{N}_C$ | $\eta_C$ [%] | $\widehat{\eta}'_C$ [%] |
|---|---|---|---|---|---|---|---|---|---|---|---|---|
| 50 | $\{5, 45\}$ | $-25.0$ | $-21.85$ | $-25.0$ | $-25.1$ | 5.31 | 5.51 | 5.54 | 5.54 | 4 | 96 | 8.33 |
| 50 | $\{4, 5, 45, 46\}$ | $-25.0$ | $-19.99$ | $-25.0$ | $-25.0$ | 5.31 | 5.72 | 5.78 | 5.78 | 7 | 92 | 15.22 |
| 50 | $\{3, 4, 5, 45, 46, 47\}$ | $-25.0$ | $-19.23$ | $-25.0$ | $-25.0$ | 5.31 | 5.92 | 6.05 | 6.05 | 21 | 88 | 47.73 |
| 50 | $\{2, 3, 4, 5, 45, 46, 47, 48\}$ | $-25.0$ | $-19.51$ | $-25.0$ | $-25.0$ | 5.31 | 6.08 | 6.35 | 6.35 | 37 | 84 | 88.10 |
| 100 | $\{9, 10, 90, 91\}$ | $-25.0$ | $-21.83$ | $-25.0$ | $-25.0$ | 2.63 | 2.73 | 2.74 | 2.74 | 4 | 96 | 4.17 |
| 100 | $\{7, 8, 9, 10, 90, 91, 92, 93\}$ | $-25.0$ | $-20.03$ | $-25.0$ | $-25.0$ | 2.63 | 2.83 | 2.86 | 2.86 | 5 | 92 | 5.43 |
| 100 | $\{5, 6, 7, 8, 9, 10, 90, 91, 92, 93, 94, 95\}$ | $-25.0$ | $-19.23$ | $-25.0$ | $-25.0$ | 2.63 | 2.93 | 2.99 | 2.99 | 13 | 88 | 14.77 |
| 100 | $\{3, 4, 5, 6, 7, 8, 9, 10, 90, 91, 92, 93, 94, 95, 96, 97\}$ | $-25.0$ | $-19.53$ | $-25.0$ | $-25.0$ | 2.63 | 3.00 | 3.14 | 3.14 | 14 | 84 | 16.67 |

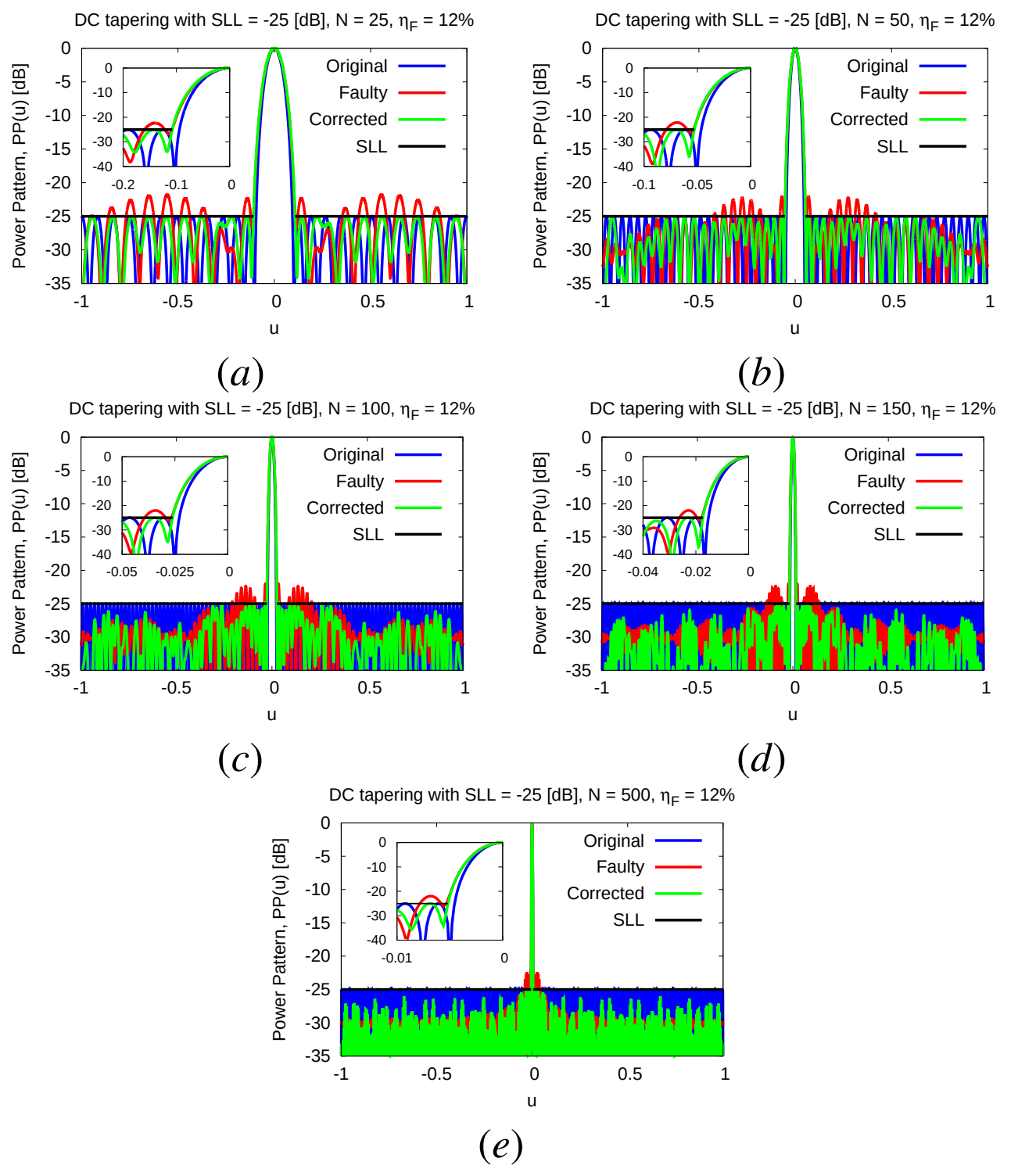

Figure 12. *Performance Analysis* ($\eta_F = 12$ %; *DC* $SLL = -25$ [dB]) - Plots of the power patterns of the original, the faulty, and the corrected arrays when (*a*) $N = 25$, (*b*) $N = 50$, (*c*) $N = 100$, (*d*) $N = 150$, and (*e*) $N = 500$.

As it can be observed in Figs. 9(*a*)-9(*b*), the number of corrections increases ($\eta_C$) with the failure rate, $\eta_F$, but more significantly for the medium array ($N_1$-elements) than for the larger one ($N_2$-elements). As a matter of fact, $0 \leq \eta_C\rfloor_{N_1} \leq 40$ %, while $0 \leq \eta_C\rfloor_{N_2} \leq 20$ %. Such an outcome is made more evident in Fig. 9(*c*) where the behavior of $\widehat{\eta}'_C$ versus the failure rate $\eta_F$ for the two array sizes is shown. It turns out that, for the medium array, more than 80 % of the $N_C$ reconfigurable elements undergo corrections when $\eta_F = 16$ %, while less than 20 % are needed when the array is $N_2$-elements wide. As expected, the larger the array the easier is the correction provided to have at disposal a suitable correction procedure able to deal with high-dimensional solution spaces in a reasonable amount of time. As for this latter item, the *CP*-method needs at most $\Delta t = 25.5$ [min] for solving the correction problems resumed in Tab. VI. For completeness, Figure 10 gives some insights on the iterative process for the synthesis of the *CP*-corrected array by showing the evolution of the backtracking vectors [Fig. 4(*b*)] for different failure percentages, $\eta_F$, but always referring to the arrangement with $N_2$ elements.

The next example is aimed at assessing the trade-off between the performance of the corrected array and the number of corrections required. The behaviour of $\widehat{SLL}$ as a function $\widehat{N}_C$ when dealing with the previous $N = 50$, $SLL = -25$ [dB], $\eta_F = 16\%$ benchmark configuration shows that (*i*) few corrections are sufficient to considerably reduce the recovered SLL (i.e., $\widetilde{SLL} - \widehat{SLL}\Big\rfloor_{\widehat{N}_C=1} \approx 2.89$ [dB] - Fig. 11), and that (*ii*) the required number of corrections decreases rapidly if slightly lower radiation performance are accepted (e.g., $\widehat{SLL}\Big\rfloor_{\widehat{N}_C=11} \approx -24.5$ [dB] vs. $\widehat{SLL}\Big\rfloor_{\widehat{N}_C=1} \approx -22.4$ [dB] - Fig. 11).

The last test case deals with the effectiveness of the proposed approach when varying the array size. Towards this end, the number of elements of an array affording a Dolph-Chebyshev pattern with $SLL = -25$ [dB] has been changed within the set of values $N \in \{25, 50, 100, 150, 500\}$. Additionally, the analysis has been repeated for different failure rates (i.e., $\eta_F \in \{4, 8, 12\}$ %) with the locations of the faulty elements as in Tab. VII. Figure 12 plots the pattern of the corrected array along with those of the original array and of the failed one when $\eta_F = 12$ %. Whatever the dimension of the array [$N = 25$ - Fig. 12(*a*); $N = 50$ - Fig.12(*b*); $N = 100$ - Fig.12(*c*); $N = 150$ - Fig. 12(*d*); $N = 500$ - Fig. 12(*e*)], the *CP* correction method successfully recovers the original *SLL* and it fits the beamwidth target, $\text{BW}_{target}$, which is slightly wider than that of the original array, but always smaller than that of the damaged array (see the insets in Fig. 12 and Tab. VII), with a limited number of corrections, which is always below the $\widehat{\eta}'_C = 60$ % of the admissible set of reconfigurable elements (Fig. 13). As expected, the percentage of corrections grows with the failure rate $\eta_F$, but, coherently with the results in Fig. 9($c$), it decreases as the array size, $N$, increases since more degrees-of-freedom are available for the correction.

Table VII
*Performance Analysis* ($DC\ SLL = -25$ [dB]) - Test cases descriptors and performance indexes.

| $N$ | *Faulty Element Indexes* | $SLL$ | $\widetilde{SLL}$ | $SLL_{target}$ | $SLL^{CP}$ | $BW$ | $\widetilde{BW}$ | $BW_{target}$ | $BW^{CP}$ | $\widehat{N}_C$ | $\eta_C$ [%] | $\widehat{\eta}'_C$ [%] |
|---|---|---|---|---|---|---|---|---|---|---|---|---|
| 25 | $\{2\}$ | $-25.0$ | $-22.28$ | $-25.0$ | $-25.0$ | 10.83 | 11.15 | 11.3 | 11.3 | 6 | 96 | 25.00 |
| 25 | $\{2, 5\}$ | $-25.0$ | $-19.27$ | $-25.0$ | $-25.0$ | 10.83 | 11.69 | 11.8 | 11.8 | 12 | 92 | 52.17 |
| 25 | $\{1, 2, 5\}$ | $-25.0$ | $-21.83$ | $-25.0$ | $-25.0$ | 10.83 | 12.43 | 12.4 | 12.4 | 13 | 88 | 59.09 |
| 50 | $\{3, 4\}$ | $-25.0$ | $-22.69$ | $-25.0$ | $-25.0$ | 5.31 | 5.47 | 5.54 | 5.54 | 4 | 96 | 8.33 |
| 50 | $\{3, 4, 9, 10\}$ | $-25.0$ | $-19.73$ | $-25.0$ | $-25.1$ | 5.31 | 5.74 | 5.78 | 5.78 | 7 | 92 | 15.22 |
| 50 | $\{1, 2, 3, 4, 9, 10\}$ | $-25.0$ | $-22.14$ | $-25.0$ | $-25.0$ | 5.31 | 6.15 | 6.05 | 6.05 | 8 | 88 | 18.18 |
| 100 | $\{5, 6, 7, 8\}$ | $-25.0$ | $-22.66$ | $-25.0$ | $-25.0$ | 2.63 | 2.71 | 2.74 | 2.74 | 3 | 96 | 3.12 |
| 100 | $\{5, 6, 7, 8, 17, 18, 19, 20\}$ | $-25.0$ | $-19.82$ | $-25.0$ | $-25.1$ | 2.63 | 2.84 | 2.86 | 2.86 | 6 | 92 | 6.52 |
| 100 | $\{1, 2, 3, 4, 5, 6, 7, 8, 17, 18, 19, 20\}$ | $-25.0$ | $-22.03$ | $-25.0$ | $-25.0$ | 2.63 | 3.06 | 2.99 | 2.99 | 6 | 88 | 6.82 |
| 150 | $\{7, 8, 9, 10, 11, 12\}$ | $-25.0$ | $-22.65$ | $-25.0$ | $-25.1$ | 1.75 | 1.80 | 1.82 | 1.82 | 3 | 96 | 2.08 |
| 150 | $\{7, 8, 9, 10, 11, 12, 25, 26, 27, 28, 39, 30\}$ | $-25.0$ | $-19.77$ | $-25.0$ | $-25.0$ | 1.75 | 1.89 | 1.90 | 1.90 | 5 | 92 | 3.62 |
| 150 | $\{1, 2, 3, 4, 5, 6, 7, 8, 9, 10, 11, 12, 25, 26, 27, 28, 29, 30\}$ | $-25.0$ | $-21.99$ | $-25.0$ | $-25.0$ | 1.75 | 2.04 | 1.99 | 1.99 | 7 | 88 | 5.30 |
| 500 | $[21, 40]$ | $-25.0$ | $-22.65$ | $-25.0$ | $-25.0$ | 0.52 | 0.54 | 0.54 | 0.54 | 2 | 96 | 0.42 |
| 500 | $[21, 40] \cup [81, 100]$ | $-25.0$ | $-19.70$ | $-25.0$ | $-25.0$ | 0.52 | 0.56 | 0.57 | 0.57 | 5 | 92 | 1.09 |
| 500 | $[1, 20] \cup [21, 40] \cup [81, 100]$ | $-25.0$ | $-21.94$ | $-25.0$ | $-25.0$ | 0.52 | 0.61 | 0.59 | 0.59 | 5 | 88 | 1.14 |

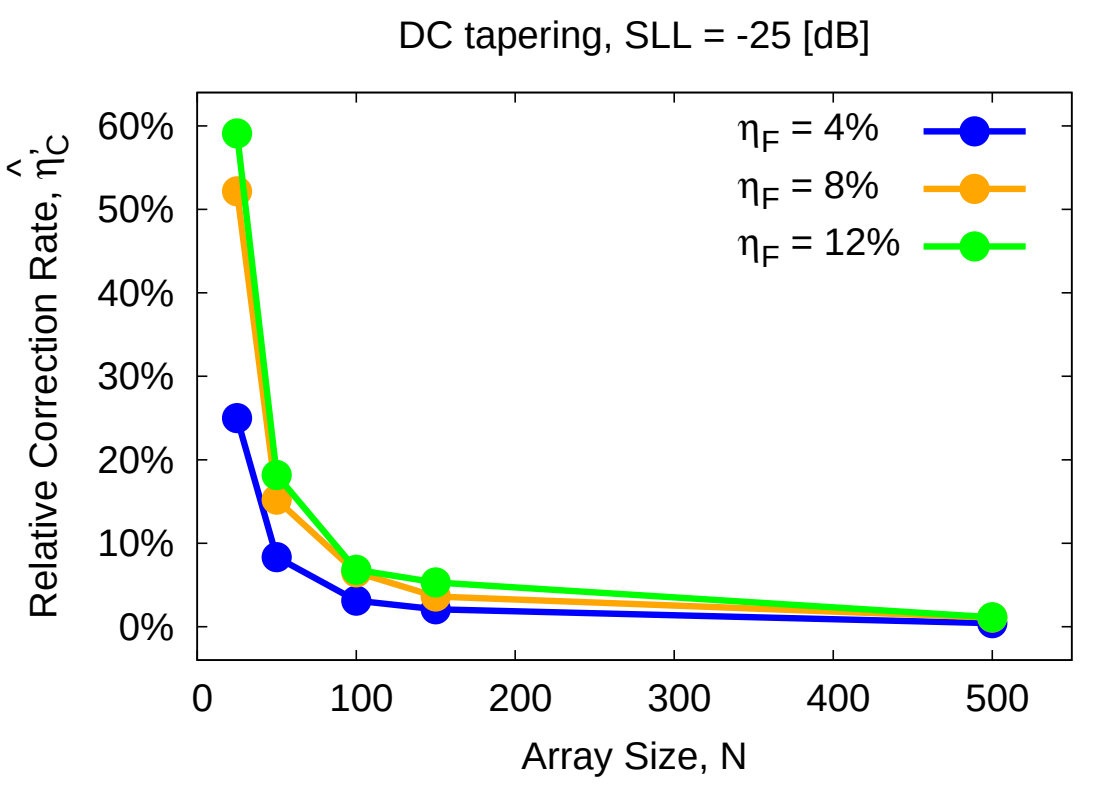

Figure 13. *Performance Analysis* ($DC\ SLL = -25$ [dB]) - Behavior of the relative correction rate, $\widehat{\eta}'_C$, versus $N$ for different values of the failure rate ($\eta_F \in \{4,\ 8,\ 12\}$ %).

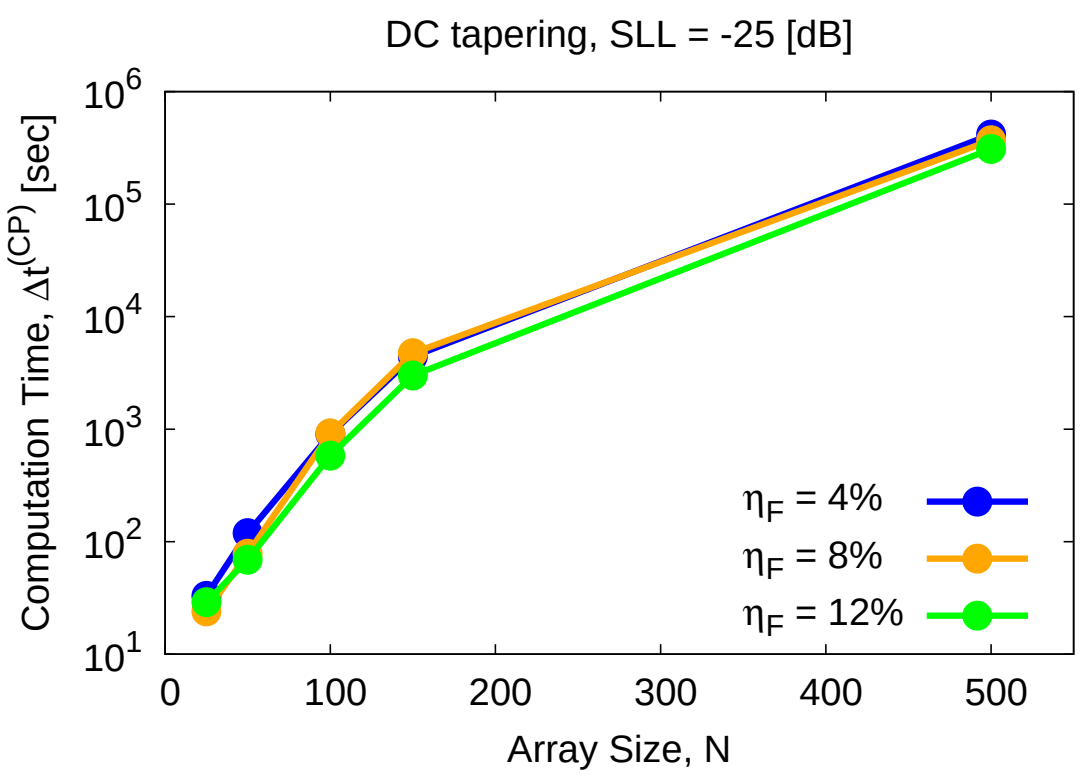

Figure 14. *Performance Analysis* ($DC\ SLL = -25$ [dB]) - *CP* method computation times, $\Delta t^{(CP)}$, as a function of the array size, $N$.

As regards the numerical efficiency of the proposed approach when large numbers of elements are at hand, the plot of $\Delta t^{(CP)}$ obtained in these latter examples (Fig. 14) show that, regardless of $\eta_F$, the computation time increases polynomially with $N$ even though the *MCFC* solution space size grows exponentially (Fig. 14). This latter result points out the effectiveness and the efficiency (e.g., $\left.\Delta t_{\min}^{(CP)}\right]_{\eta_F=8\%} = 24.0$ [s], $\left.\Delta t_{max}^{(CP)}\right]_{\eta_F=8\%} = 3.69\times10^5$ [sec] - Fig. 14) of the *CP* method despite its non-optimized software implementation, which turns out reliable even when dealing with high-dimensional *MCFC* problems including hundreds of array elements (e.g., $N = 500$ - Fig. 14). Such a conclusion is also consistent with the well-known efficiency of *CP*-based approaches in electromagnetics [31], which usually outperform bare evolutionary optimization techniques (such as *GA* [21][23]) in terms of solution speed thanks to their capability to effectively leverage on the *a-priori* knowledge on the solution sparsity [31].

## V. Conclusions and Final Remarks

A method to address the *MCFC* problem in linear arrays has been presented. Within the *CP* framework, such an approach integrates a $\ell_1$-norm relaxation of the $\ell_0$-norm with a backtracking strategy to synthesize the minimum number of corrections of the damaged array for recovering the original pattern features in a limited amount of time when dealing with large apertures, as well. The proposed correction method has been assessed in different scenarios and comparisons with reference state-of-the-art techniques have been performed.

The numerical assessment has shown that

- the *CP* method positively compares with competitive alternatives in terms of both number of corrections and recovered pattern features and/or fitted user-requirements;
- the number of required corrections is lower than that of the reconfigurable array elements and, often, it is smaller than the number of failures;
- the *CP* method faithfully, reliably, and efficiently performs in correcting small, medium, and large array with hundreds of elements.

Thanks to its effectiveness in dealing with high-dimension solution spaces, future research studies - beyond the scope of the present paper - will consider the extension of the *CP* correction method to two- and three-dimensional arrays.

## Acknowledgments

A. Massa wishes to thank E. Vico for her never-ending inspiration, support, guidance, and help.